\documentclass[manuscript, screen, acmsmall]{acmart}
\AtBeginDocument{%
  }

\setcopyright{acmlicensed}
\copyrightyear{2026}
\acmYear{2026}
\acmDOI{XXXXXXX.XXXXXXX}

\acmJournal{HEALTH}
\acmVolume{1}
\acmNumber{1}
\acmArticle{1}
\acmMonth{1}

\usepackage[capitalize]{cleveref}
\usepackage{todonotes}
\usepackage{fontawesome}
\usepackage{wrapfig}
\usepackage{opensans}
\newboolean{SHOWCOMMENTS}
\setboolean{SHOWCOMMENTS}{true}
\ifthenelse{\boolean{SHOWCOMMENTS}}{
\newcommand{\tdMK}[1]
{\todo[color=blue!25,inline]{\footnotesize{\textbf{ Martin:}} #1}}
}{
\newcommand{\tdMK}[1]{}
}

\ifthenelse{\boolean{SHOWCOMMENTS}}{
\newcommand{\tdHZ}[1]
{\todo[color=green!25,inline]{\footnotesize{\bf Henrik:} #1}}
}{
\newcommand{\tdHZ}[1]{}
}

\ifthenelse{\boolean{SHOWCOMMENTS}}{
\newcommand{\tdSM}[1]
{\todo[color=orange!25,inline]{\footnotesize{\bf Shahbaz:} #1}}
}{
\newcommand{\tdSM}[1]{}
}

\ifthenelse{\boolean{SHOWCOMMENTS}}{
\newcommand{\tdAW}[1]
{\todo[color=teal!25,inline]{\footnotesize{\bf Anna:} #1}}
}{
\newcommand{\tdAW}[1]{}
}

\ifthenelse{\boolean{SHOWCOMMENTS}}{
\newcommand{\tdFJ}[1]
{\todo[color=yellow!25,inline]{\footnotesize{\bf Fredrik:} #1}}
}{
\newcommand{\tdFJ}[1]{}
}

\ifthenelse{\boolean{SHOWCOMMENTS}}{
  
  \newcommand{\tdCC}[1]
  {\todo[color=red!10,inline]{\footnotesize{\textbf{ Cas:}} #1}}
}{
  \newcommand{\tdCC}[1]{}
}

\ifthenelse{\boolean{SHOWCOMMENTS}}{
\newcommand{\tdMJ}[1]
{\todo[color=orange!25,inline]{\footnotesize{\bf Mariama:} #1}}
}{
\newcommand{\tdMJ}[1]{}
}

\ifthenelse{\boolean{SHOWCOMMENTS}}{
\newcommand{\tdDK}[1]
{\todo[color=red!25,inline]{\footnotesize{\textbf{ David:}} #1}}
}{
\newcommand{\tdDK}[1]{}
}

\renewcommand{\d}[1]{\mathrm{d}#1}

\newcommand{\reviewanonymized}[2]{%
  \ifx\vismode\anonymized
    #1%
  \else
    #2%
  \fi
}

\usepackage{graphicx} 
\usepackage{subcaption}

\usepackage{tikz}
\usetikzlibrary{shapes.multipart} 
\usetikzlibrary{positioning,decorations.pathreplacing}

\setcopyright{none}
\begin{document}

\title[A full software stack for epidemic disease management]{A full software stack for epidemic disease management: Unlocking the joint potential of software technology and supercomputing}

\reviewanonymized{
\author{Jonas Gilg}
\email{jonas.gilg@dlr.de}
\orcid{0000-0001-9130-0778}
\authornote{Shared first authors. These authors contributed equally to this work.}
\affiliation{
  \institution{Institute of Software Technology, German Aerospace Center}
  \city{Brunswick}
  \country{Germany}
}

\author{Johann Fredrik Jadebeck}
\email{j.jadebeck@fz-juelich.de}
\orcid{0000-0002-5026-1546}
\authornotemark[1]
\affiliation{%
  \institution{Institute of Bio- and Geosciences, IBG-1: Biotechnology, Forschungszentrum J\"ulich GmbH}
  \city{Jülich}
  \country{Germany}
}
\affiliation{%
  \institution{Computational Systems Biotechnology (AVT.CSB), RWTH Aachen University}
  \city{Aachen}
  \country{Germany}
}

\author{Mariama Jaiteh}
\email{mariama.jaiteh@helmholtz-hzi.de}
\orcid{0000-0002-9229-5314}
\authornotemark[1]
\affiliation{
  \institution{Helmholtz Centre for Infection Research}
  \city{Brunswick}
  \state{Lower Saxony}
  \country{Germany}
}

\author{David Kerkmann}
\email{david.kerkmann@theoretical-biology.de}
\orcid{0009-0007-9109-096X}
\authornotemark[1]
\affiliation{
  \institution{Helmholtz Centre for Infection Research}
  \city{Brunswick}
  \state{Lower Saxony}
  \country{Germany}
}

\author{Niklas Medinger}
\orcid{0009-0006-0494-6137}
\email{niklas.medinger@cispa.de}
\authornotemark[1]
\affiliation{%
\institution{CISPA Helmholtz Center for Information Security}
\city{Saarbr\"{u}cken}
\country{Germany}
}

\author{Shahbaz Memon}
\email{m.memon@fz-juelich.de}
\orcid{0000-0002-5229-5774}
\authornotemark[1]
\affiliation{%
  \institution{J\"ulich Supercomputing Centre, Forschungszentrum J\"ulich GmbH}
  \city{Jülich}
  \country{Germany}
}

\author{Anna Wendler}
\email{anna.wendler@dlr.de}
\orcid{0000-0002-1816-8907}
\authornotemark[1]
\affiliation{%
  \institution{Institute of Software Technology, German Aerospace Center}
  \city{Cologne}
  \country{Germany}
}

\author{Moritz Zeumer}
\email{moritz.zeumer@dlr.de}
\orcid{0000-0001-5717-1554}
\authornotemark[1]
\affiliation{
  \institution{Institute of Software Technology, German Aerospace Center}
  \city{Brunswick}
  \country{Germany}
}

\author{Henrik Zunker}
\email{henrik.zunker@dlr.de}
\orcid{0000-0002-9825-365X}
\authornotemark[1]
\affiliation{%
  \institution{Institute of Software Technology, German Aerospace Center}
  \city{Cologne}
  \country{Germany}
}

\author{Maximilian Betz}
\email{m.betz@fz-juelich.de}
\orcid{0009-0001-9181-9999}
\affiliation{%
  \institution{Institute of Climate and Energy Systems, ICE-1: Energy Systems Engineering, Forschungszentrum J\"ulich GmbH}
  \city{Jülich}
  \country{Germany}
}

\author{Ralf Hannemann-Tamas}
\email{ralf.ht@gmail.com}
\orcid{0000-0001-9765-3013}
\affiliation{%
  \institution{Institute of Climate and Energy Systems, ICE-1: Energy Systems Engineering, Forschungszentrum J\"ulich GmbH}
  \city{Jülich}
  \country{Germany}
}

\author{Jonas Immanuel Heinicke}
\orcid{0000-0002-8442-3187}
\affiliation{
  \institution{Helmholtz Centre for Infection Research}
  \city{Brunswick}
  \state{Lower Saxony}
  \country{Germany}
}

\author{Julian Litz}
\email{julian.litz@icloud.com}
\orcid{0009-0009-9883-4786}
\affiliation{%
  \institution{Institute of Climate and Energy Systems, ICE-1: Energy Systems Engineering, Forschungszentrum J\"ulich GmbH}
  \city{Jülich}
  \country{Germany}
}

\author{Achim Basermann}
\affiliation{%
  \institution{Institute of Software Technology, German Aerospace Center, Cologne}
  \city{Cologne}
  \country{Germany}
}
\orcid{0000-0003-3637-3231}
\email{achim.basermann@dlr.de}

\author{Cas Cremers}
\affiliation{%
\institution{CISPA Helmholtz Center for Information Security}
\city{Saarbr\"{u}cken}
\country{Germany}
}
\orcid{0000-0003-0322-2293}
\email{cremers@cispa.de}

\author{Manuel Dahmen}
\email{m.dahmen@fz-juelich.de}
\orcid{0000-0003-2757-5253}
\affiliation{%
  \institution{Institute of Climate and Energy Systems, ICE-1: Energy Systems Engineering, Forschungszentrum J\"ulich GmbH}
  \city{Jülich}
  \country{Germany}
}

\author{Andreas Gerndt}
\email{andreas.gerndt@dlr.de}
\orcid{0000-0002-0409-8573}
\affiliation{%
  \institution{Institute of Software Technology, German Aerospace Center, Brunswick and Center for Industrial Mathematics, University of Bremen}
  \city{Bremen}
  \country{Germany}
}

\author{Jens Henrik G\"obbert}
\email{j.goebbert@fz-juelich.de}
\orcid{0000-0002-3807-6137}
\affiliation{%
  \institution{J\"ulich Supercomputing Centre, Forschungszentrum J\"ulich GmbH}
  \city{Jülich}
  \country{Germany}
}
\author{Bj\"orn Hagemeier}
\email{b.hagemeier@fz-juelich.de}
\orcid{0000-0003-1528-0933}
\affiliation{%
  \institution{J\"ulich Supercomputing Centre, Forschungszentrum J\"ulich GmbH}
  \city{Jülich}
  \country{Germany}
}

\author{Carolina J. Klett-Tammen}
\affiliation{%
  \institution{Helmholtz Centre for Infection Research}
  \city{Braunschweig}
  \country{Germany}
}
\orcid{0000-0001-9685-5369}
\email{Carolina.Klett-Tammen@helmholtz-hzi.de}

\author{Berit Lange}
\affiliation{%
  \institution{Helmholtz Centre for Infection Research}
  \city{Braunschweig}
  \country{Germany}
}
\orcid{0000-0002-9325-9307}
\email{Berit.Lange@helmholtz-hzi.de}

\author{Katharina N\"oh}
\email{k.noeh@fz-juelich.de}
\orcid{0000-0002-5407-2275}
\affiliation{%
  \institution{Institute of Bio- and Geosciences, IBG-1: Biotechnology, Forschungszentrum J\"ulich GmbH}
  \city{Jülich}
  \country{Germany}
}

\author{Sarah Straßburger}
\authornote{Shared corresponding authors.}
\affiliation{%
  \institution{Helmholtz Centre for Infection Research}
  \city{Braunschweig}
  \country{Germany}
}
\orcid{0009-0003-1353-0000}
\email{sarah.strassburger@helmholtz-hzi.de}

\author{Michael Meyer-Hermann}
\authornotemark[2]
\affiliation{%
  \institution{Department of Systems Immunology and Braunschweig Integrated Centre of Systems Biology, Helmholtz Centre for Infection Research, Braunschweig;}
  \institution{Institute for Biochemistry, Biotechnology and Bioinformatics, 
Technische Universität Braunschweig, Braunschweig;}
  \institution{and Lower Saxony Center for Artificial Intelligence and Causal Methods in Medicine (CAIMed), Hannover.}
  \country{Germany}
}
\orcid{0000-0002-4300-2474}
\email{mmh@theoretical-biology.de}

\author{Martin J. Kühn}
\authornotemark[2]
\affiliation{%
  \institution{Institute of Software Technology, German Aerospace Center, Cologne;}
  \institution{and Bonn Center for Mathematical Life Sciences and Life and Medical Sciences Institute, University of Bonn}
  \city{Bonn}
  \country{Germany}
}
\orcid{0000-0002-0906-6984}
\email{martin.kuehn@dlr.de}
}
{
}

\renewcommand{\shortauthors}{\reviewanonymized{Gilg, Jadebeck, Jaiteh, Kerkmann, Medinger, Memon, Wendler, Zeumer, Zunker et al.}{}}

\begin{abstract}
Infectious diseases remain a major threat to human societies. During the recent COVID-19 pandemic, mathematical modeling and extensive computer simulations proved highly effective in supporting public health experts and decision makers.

Despite these advances, the full potential of modern modeling approaches and digital technologies has not yet been realized. Many critical tasks -- including expert consultations, model execution, scenario analyses, report preparation, and result communication -- still relied heavily on manual, human-driven processes with each manual interaction introducing avoidable delays and limiting responsiveness during rapidly evolving outbreaks.

Pandemic preparedness should opt for automated workflows and seamlessly integrated software modules that can improve pandemic mitigation capabilities by substantially reducing response times. For this step, we require robust and flexible computational infrastructure capable of supporting heterogeneous hardware and continuously evolving infectious-disease models. In addition, data sources need to be dynamically integrated. Managing such demands needs infrastructure that supports automated high-performance computing (HPC) workflows. Beyond computational performance, software infrastructure must ensure secure user and data management to comply with data-protection regulations and provide clear, transparent presentation of results to both decision makers and the public. 

Meeting the aforementioned challenges requires tight integration of state-of-the-art scientific software with modern, scalable infrastructure that can leverage supercomputing resources when necessary. For rapid deployment in future epidemic or pandemic scenarios, adherence to the FAIR principles for research software is critical to ensure reusability and sustainability.
\end{abstract}





\maketitle

\section{Introduction}


Human societies are largely impacted as soon as a novel major outbreak of infectious diseases occurs. Mathematical modeling and computer simulations have been highly effective in supporting public health experts and decision makers in the recent COVID-19 pandemic~\cite{european_centre_for_disease_prevention_and_control_public_2024}. Furthermore, automated data pipelines~\cite{10.1093/haschl/qxad080} and digital contact tracing~\cite{ferretti2020quantifying} were instrumental in supporting the timely response to and mitigation of COVID-19 spread.

However, the vast potential of today's models and technology have not yet been fully utilized. Requests for expertise, predictions and scenario analyses, preparation of reports, communication of results, or the provision of model simulations were often based on human-to-human interaction and manual interactions with software components. While the generated knowledge already drastically improved the capabilities of decision makers and public health experts to fight COVID-19, substantial and unnecessary delays are the result of each manual interaction. 
To further enhance pandemic preparedness, robust and flexible computational infrastructure is required across both heterogeneous hardware and evolving infectious-disease models. 
For instance, during the COVID-19 pandemic, novel data was dynamically integrated and complex expert models were continuously developed~\cite{SHATTOCK2022100535,OpenABM,adhikari2020,kerr_covasim_2021,mueller2021episim,vermeulen2021,zunker_novel_2024,WENDLER2026129636,bicker_hybrid_2025,PLOTZKE2026823,KERKMANN2025110269,bicker2026memilio}.
To handle computationally costly models, any infrastructure also needs to support automated high-performance computing (HPC).
The targeted infrastructure must also provide robust user and data management to ensure compliance with data-protection regulations as well as clear presentation of results to both decision makers and the public. 
Meeting these requirements necessitates tight integration of state-of-the-art scientific software with modern, robust infrastructure that leverages supercomputing resources where required. For rapid deployment in future epidemic or pandemic settings, reproducibility and re-usability are key and it is thus important to closely align to the FAIR principles for research software (FAIR4RS)~\cite{barker_introducing_2022}.

In this paper, we introduce our open-source software stack for epidemic data integration, modeling, and interactive feedback to support public health decision making and information dissemination during epidemic times. The platform is the culmination of an interdisciplinary project consortium that combines expertise from various domains such as epidemiology, immunology, biology, public health, mathematics, computer science, high-performance computing, and visualization. Using Germany as a case study, we conducted a multi-year pilot phase of our platform, demonstrating the versatility and utility for epidemic mitigation and evaluating how it would have supported the daily work of local health authorities (LHA). With all software components integrated, the platform provides decision makers and public health experts with (immediate) feedback on requests for situation assessments and alternative scenarios, visualizing the results based on daily integrated data and also enabling the automatic integration of model updates into workflows. With the open-source provision of the elementary software with open licenses, our software stack is easily reusable, in its entirety or in parts, and extensible for upcoming epidemic challenges.

Our paper is structured as follows. In~\cref{sec:related}, we provide a short overview of existing software frameworks, stacks, and platforms with respect to pandemic mitigation for public health experts and decision makers. In~\cref{sec:methods}, we provide detailed insights into our platform with respect to (i) data sources (\cref{sec:datasources}), (ii) mathematical modeling approaches (\cref{sec:modeling}), (iii) data integration (\cref{sec:dataint}), (iv) front-end and REST API (\cref{sec:frontend_restapi}), (v) authentication and user management (\cref{sec:usermanage}), (vi) workflow orchestration (\cref{sec:automated}), and eventually (vii) the connection to high-performance computing and its infrastructure (\cref{sec:hpc}). In the results section, we present (i) the platform itself (\cref{sec:results_framework}), (ii) a proof-of-concept with synthetic local data integration (\cref{sec:local_data_scenario}), (iii) the results of testing and user experience (\cref{sec:UX},, and (iv) timings and performance measures (\cref{sec:performance}). Finally, we conclude in~\cref{sec:conclusion}.

\section{Related work}\label{sec:related}

In particular with the COVID-19 pandemic, a lot of models and modeling frameworks have been proposed~\cite{SHATTOCK2022100535,OpenABM,kerr_covasim_2021,mueller2021episim,vermeulen2021,adhikari2020,bicker2026memilio}. However, these alone represent only one core element of the software stack, as presented in this paper. Based on our search on literature and software and to the best of our knowledge, there is no fully integrated and automated platform for pandemic response. 

Analyzing the existing landscape, insights were drawn from a prospectively registered scoping review (Open Science Framework; OSF ID: q32kg), which systematically maps global efforts to develop infectious disease forecasting tools for public-health decision-making~\cite{fischer_global_2025}.

A substantial proportion of identified tools cover only isolated elements of a forecasting workflow, most commonly in the form of static dashboards or disease-specific visualizations without scope for regional adaptation or scenario exploration~\cite{dong_interactive_2020,who_dashboard,hasell_cross-country_2020}. Additionally, several platforms provide model outputs without interactive parameterization or integration of contextual inputs, limiting their suitability for operational decision-making in local or subnational public-health settings~\cite{desai_real-time_2019}. Furthermore, many previously described tools are discontinued, not maintained, or no longer accessible following the termination of project-based funding~\cite{reiner_modeling_2021,panovska-griffiths_determining_2020,funk_real-time_2018}, constraining their utility for sustained preparedness or routine public-health operations.

These findings underscore relevant challenges in translating forecasting research into practice and align with the design objectives of platforms such as presented here. In particular, our initiatives focus on regionally adaptable forecasting capabilities, the integration of locally contextualized inputs, and the provision of an interactive, user-oriented visual analytics environment intended to support real-world decision-making. By combining flexible modeling components, locally relevant data structures, and operational usability within a single platform architecture, we seek to address the fragmented landscape identified in the scoping review and to deliver a forecasting solution that is implementation-ready. A comprehensive synthesis of the scoping review's findings will be reported in a dedicated publication~\cite{fischer_global_2025}.


\section{Approach and methods}\label{sec:methods}

The target users of our platform, ranging from LHAs to decision makers and the general public need to have swift and clear access to ongoing epidemic dynamics with potential outcomes for alternative scenarios to better manage or understand mitigation efforts. First, this creates the need for advanced mathematical modeling that is based on sound data and data integration. Secondly, we need well defined software interfaces, automated workflows, and HPC integration to realize a swift and seamlessly integrated software stack. This section outlines the details of data sources and data integration together with the methodological foundation of our platform, covering modeling and computation, data pipelines, visualization, and communication of results to decision makers.


\subsection{Data sources}\label{sec:datasources}

\begin{figure}
    \centering
    \includegraphics[width=0.8\linewidth]{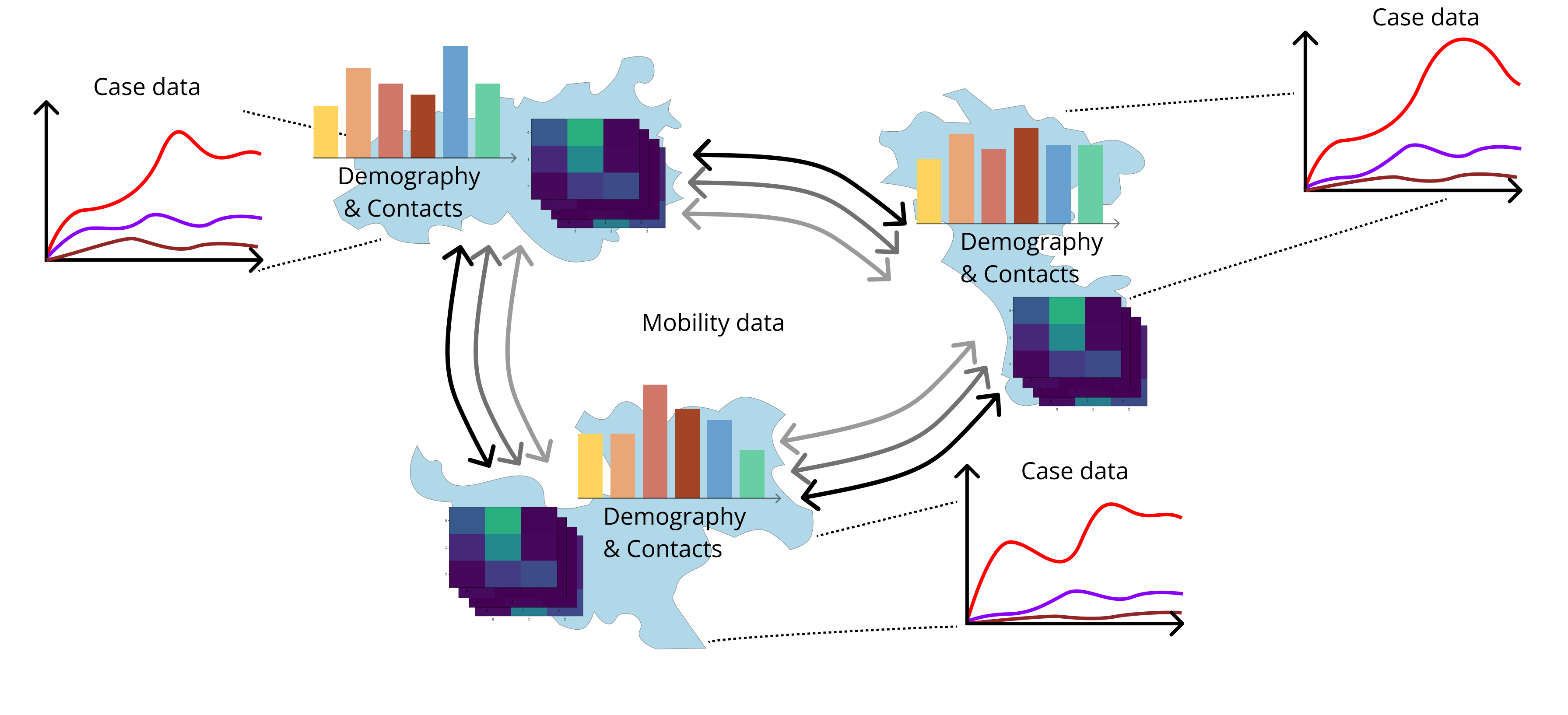}
    \caption{\textbf{A schematic overview of spatially and demographically resolved data for epidemic mitigation.}}
    \label{fig:data}
\end{figure}

Reliable data is a key element for mitigating infectious diseases. In the following, we describe the data sources available in the recent COVID-19 pandemic and which we expect to be available in future pandemic contexts. While often the type of data is discussed, the (potential) stratification of the data is also a key element discussed and considered here. As the last meta-dimension of data, we expect data to be provided on a daily basis and with minimal delay to act with a most up-to-date information basis. For a schematic view of the data; see~\cref{fig:data}.

In an epidemic setting, we expect to obtain daily reported numbers of positive confirmed, severe, and critical cases, i.e., hospitalized and intensive care cases, as well as deaths. Since a novel pathogen such as SARS-CoV-2 is unlikely to pose an equal threat to individuals of all age groups, the data should be stratified by age. For our SARS-CoV-2 demonstrator case in Germany, confirmed cases and deaths were reported with six age groups of 0-4, 5-14, 15-34, 35-39, 60-79, and 80+ years old~\cite{robert_koch_institut_2025_17627234}, and hospitalizations and intensive care numbers without age stratification~\cite{robert_koch_institut_2025_17623109}.

Eventually, the spread of SARS-CoV-2 has often been heterogeneous on a spatial scale and data was reported on district level. For the 400 districts we then included demographic data~\cite{statistisches_bundesamt_destatis_population_nodate} that was aggregated on the age groups specified above. To accurately address spatiotemporal heterogeneity, our models (see~\cref{sec:modeling}) require mobility data or an estimate of regular mobility exchanges between all districts. Here, we used data extrapolated from~\cite{bmas_pendlerverflechtungen_2020} as well as a data set for inter-district commuting based on a macroscopic transport model and a synthetic population for Germany~\cite{winkler_impact_2020}. In order to quantify the daily number of contacts, we used~\cite{MHJ08} and its projections~\cite{PCJ17}, split up into the categories ``Home'', ``School'', ``Work'', and ``Other''. For the school contacts, we additionally combined~\cite{FAM12}. 

During the first vaccination phases in 2021 and 2022, we additionally used the database~\cite{robert_koch_institut_2024_12697471}. As data was reported with the location of vaccination, we approximated the local vaccination ratios as given by~\cite{koslow_appropriate_2022}. However, since no vaccination registry is available for Germany, immunity data could not be estimated once a substantial part of the population had recovered or was vaccinated a third or fourth time. In later phases~\cite{zunker_novel_2024}, three different immunity layers were assessed through seropositivity studies~\cite{lange_berit_2022_6968574}.

General parameters such as the probability of infection, the time to stay infected, the proportion of asymptomatic, severe, and critical infections were quantified from~\cite{kuhn_assessment_2021,koslow_appropriate_2022} and the references therein. 

While the case data described above was provided with short delays, our platform also integrates an option, for each LHA, to supplement the public data sources with own, local data sets. These local data sets may contain additional information regarding the individual vaccination history or virus variant to allow for more precise initialization of models.

\subsection{Mathematical modeling}\label{sec:modeling}

Our modeling components build on the MEmilio framework~\cite{Bicker_MEmilio_v2_1_0,bicker2026memilio} which is not tied to a single disease, and where models can be used or easily adapted for different respiratory diseases using already existing structures. MEmilio supports models based on ordinary differential equations (ODEs) which can be extended to metapopulation settings~\cite{kuhn_assessment_2021, zunker_novel_2024, ZUNKER2025116782}. In addition, it allows for equation-based models using Erlang or Gamma~\cite{PLOTZKE2026823} or even arbitrarily distributed disease stage transition times~\cite{WENDLER2026129636}. Furthermore, agent-based models (ABMs)~\cite{KERKMANN2025110269} or hybrid agent-metapopulation-based~\cite{bicker_hybrid_2025} models are available. In this work, we focus on the metapopulation formulations and use the ABM for our proof-of-concept data generation process.

\subsubsection{Requirements}\label{sec:modeling_requirements}

To capture the dynamics of a specific pathogen, it is crucial to set up a model structure that accounts for the most important aspects of the corresponding disease. Therefore, we identified several general and SARS-CoV-2 specific requirements for a model inside the software stack. These are
\begin{itemize}
\item the consideration of asymptomatic development and a presymptomatic infectious phase,
\item the inclusion of age-dependent vulnerability,
\item the inclusion of heterogeneous immunity and vaccination,
\item the inclusion of severe, critical, and deceased states,
\item the inclusion of spatial resolution,
\item the possibility to compute different scenarios,
\item the efficient and optimized implementation of the model 
\end{itemize}
to overall allow public health experts and LHAs study different aspects and local developments of disease dynamics.

\subsubsection{Modeling approach}\label{sec:modeling_approach}
\begin{figure}
    \centering
    \includegraphics[width=1.0\linewidth]{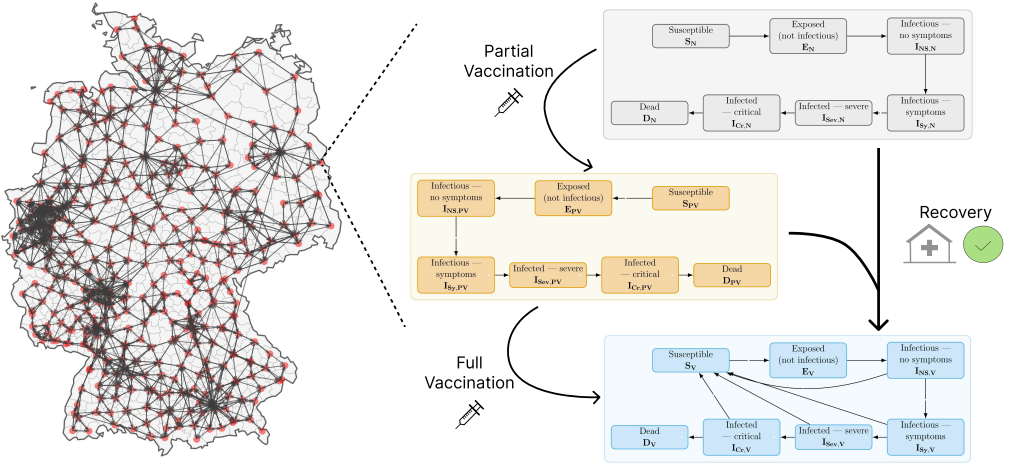}
    \caption{\textbf{Metapopulation coupling and disease progression in the \textit{SECIRVVS} model.} The left panel shows all 400 districts of Germany, with the centroid of each district represented by a red dot. The local models are connected with edges representing mobility. Based on commuter information from~\cite{bmas_pendlerverflechtungen_2020}, we visualize the 2\,500 strongest commuter links. Each district contains its own \textit{SECIRVVS} model. As shown in the flow chart on the right, each model has three immunity layers and is stratified in the compartments susceptible~($S$), exposed/non‑infectious~($E$), infectious without symptoms~($I_{NS}$), infectious with symptoms~($I_{Sy}$), hospitalized~($I_{Sev}$), intensive care~($I_{Cr}$), and dead~($D$). Arrows indicate transitions and recovery moves individuals to better‑protected layers. Vaccinations are applied as a daily discrete step between susceptible compartments.}
    \label{fig:ode_metapop}
\end{figure}

To address the aforementioned requirements, we used and tested an age‑resolved \textit{SECIR}-type metapopulation model with two levels of vaccination, denoted \textit{SECIRVVS}; see~\cref{fig:ode_metapop} for a flow chart between the disease states and~\cite{koslow_appropriate_2022} for more details. 

To make this paper self-contained, we provide a minimal introduction of the SECIRVVS model used throughout our pilot development and testing phase. The model allows for age groups $i=1,\ldots,n_A$, where we chose $n_A=6$ according to the provided data sources. Furthermore, we have spatial stratification with $k=1,\ldots,n_S$, which was set to $n_S=400$ to model Germany's 400 districts, and three vaccination or immunity layers $M\in\{\mathrm{N},\mathrm{PV},\mathrm{V}\}$. The total number of persons in age group $i\in\{1,\ldots,n_A\}$ and region $k\in\{1,\ldots,n_S\}$ is given by $N^{(k)}_{i}$ and similarly, we have the local number of individuals in the particular compartments by $S^{(k)}_{i}$, $E^{(k)}_{i}$, et cetera. The (time–dependent) contact matrix $\Phi^{(k)}(t)$ is given by its entries of intra- or inter-age group contact rates by $\phi^{(k)}_{i,j}(t)$, $i,j\in\{1,\ldots,n_A\}$. Furthermore, we include isolation and quarantining in our model. Then, the fraction of non-isolated carriers and symptomatic infected individuals is given by $\xi^{(k)}_{I_{NS},j}(t)$ and $\xi^{(k)}_{I_{Sy},j}(t)$, respectively. The force of infection for immunity layer $M$ reads
\begin{align}
\label{eq:foi}
\lambda^{(k)}_{M,i}(t)
= \rho^{(k)}_{M,i}(t)\,
\sum_{j} \phi^{(k)}_{i,j}(t)\,
\frac{\xi^{(k)}_{I_{NS},j}(t)\,\sum_{M'} I^{(k)}_{NS,M',j}(t) + \xi^{(k)}_{I_{Sy},j}(t)\,\sum_{M'} I^{(k)}_{Sy,M',j}(t)}{N^{(k)}_{j}(t)-\sum_{M'}D^{(k)}_{M',j}(t)}.
\end{align}
Here $\rho^{(k)}_{M,i}(t)=s_{\kappa_S}(t)\,\rho^{(k)}_{0,M,i}$ captures baseline transmissibility $\rho^{(k)}_{0,M,i}$ and seasonality $s_{\kappa_S}(t)=1+\kappa_S\sin\!\big(\pi(t/182.5+1/2)\big)$, where $\kappa_S$ is a seasonality constant that varied around 0.2. While variant shares can be incorporated multiplicatively, variant regime change was not the focus of this model. For more details; see~\cite{koslow_appropriate_2022}.

Then, the dynamics in the susceptible compartment of layer $M$ can be described by
\begin{align}
\label{eq:S-balance}
\frac{\d}{\d t} S^{(k)}_{M,i}(t)
= -\,\lambda^{(k)}_{M,i}(t)\,S^{(k)}_{M,i}(t).
\end{align}
Other transitions can be defined in a more generic way. Let $z^{(k)}_{q,M,i}$ denote the $q$-th infection state of our model for age group $i$ and region $k$ and immunity layer $M\in\{\mathrm{N},\mathrm{PV},\mathrm{V}\}$. Using mean sojourn times $T^{(k)}_{z_{q,N,i}}$ and transition probabilities $\mu^{(k)}_{z_{q+1,N,i}\,;\,z_{q,N,i}}$ to transit from (naive immunity) state $z_{q,N,i}$ to $z_{q+1,N,i}$, the dynamics in an infection state can be described by
\begin{align}
\label{eq:generic}
\frac{\d}{\d t} z^{(k)}_{q+1,N,i}(t)
= \frac{\mu^{(k)}_{z_{q+1,N,i}\,;\,z_{q,N,i}}}{T^{(k)}_{z_{q,N,i}}}\,z^{(k)}_{\,q,N,i}(t)
\;-\;\frac{z^{(k)}_{q+1,N,i}(t)}{T^{(k)}_{z_{q+1,N,i}}}.
\end{align}

Individuals with recovery from non-symptomatic, symptomatic, severe, and critical developments are routed to the better protected immunity layers and immunity layer $M\in\{PV, V\}$ modifies both risks and durations. Let $\mu^{(k)}_{\cdot,N,i;\cdot,N,i}$ be the transition probabilities for naive individuals. For risks, we use the naive transition probabilities and add reduction factors $p$ such that
\begin{align}
\label{eq:reweight}
\mu^{(k)}_{z_{q+1},M,i\,;\,z_{q},M,i}
= \frac{p_{z_{q+1},M,i}}{p_{z_{q},M,i}}\;
\mu^{(k)}_{z_{q+1},N,i\,;\,z_{q},N,i}
\end{align}
with parameters $p_{E,M,i}$, $p_{I,M,i}$, $p_{H,M,i}$, $p_{U,M,i}$, $p_{D,M,i}$ representing the age-specific improved immunity due to partial or full vaccination (and prior infection) to reflect better protection against any infection, symptoms, hospitalization, intensive care treatment, and death.
Mild infectious periods in protected layers are shortened via $T^{(k)}_{z_q,M,i}=\kappa_M\,T^{(k)}_{z_q,N,i}$ with $\kappa_M\in(0,1]$. Non-pharmaceutical interventions (NPIs) are realized through a change in the contact patterns $\Phi$ or through a reduction of mobility on the edges of the graph. Vaccinations are treated outside the dynamical system formulation as a daily update between susceptible compartments of different layers. Recovered individuals are routed to the best protected immunity layer.

To capture spatial dynamics, we embed the local \textit{SECIRVVS} models in a metapopulation approach~\cite{kuhn_assessment_2021, zunker_novel_2024} where each district is represented by its own local \textit{SECIRVVS} model. The regions are coupled via mobility edges that represent commuting and traveling activities between regions. This allows us to model the spatial spread of infections due to mobility between regions. In~\cref{fig:ode_metapop}, we visualize the $2\,500$ strongest commuter links between the districts of Germany.

Beyond the previous description, MEmilio's modular structure allows to flexibly adapt the model complexity to the specific situation in order to meet the requirements for modeling different pathogens, strains, or regions. The model from~\cite{koslow_appropriate_2022} as described above has been specifically designed for COVID-19 in Germany when detailed data on age-stratified cases, ICU occupancy and vaccination data was available. It was thus chosen as the state-of-the-art demonstrator in the described software stack.

Although ODE systems can be solved efficiently within the MEmilio framework, due to the resolution on district level and the inclusion of mobility exchange, computation cannot be done on-the-fly -- in particular when a wide range of scenarios is considered.
To support on-the-fly exploration, we provide machine learning (ML) surrogate models trained on curated simulation outcomes and which replace the mechanistic expert model. The approach based on graph neural networks has been described in~\cite{schmidt2025} and allows instantaneous feedback to public health experts while still integrating spatial and age resolution. Deployment and continuous retraining hooks are integrated in the pipeline but not enabled in production yet. 

\begin{figure}
    \centering
    \includegraphics[width=1.0\linewidth]{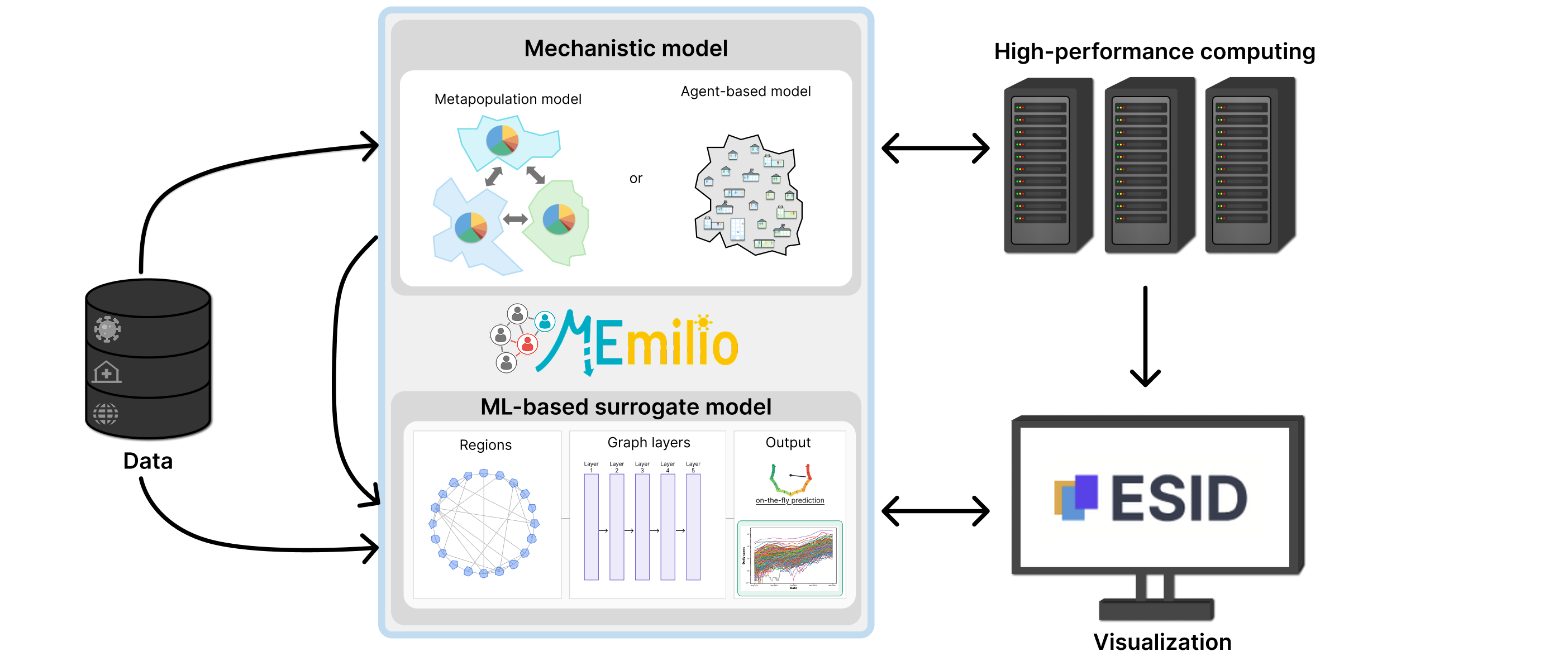}
    \caption{\textbf{Overview of mathematical models and their position within the pipeline.} MEmilio provides mechanistic models such as the graph-based SECIRVVS metapopulation model. As alternative, an ABM could be chosen for integration. Furthermore, an ML surrogate model trained on data generated by the metapopulation models allows on-the-fly simulations and could be integrated into the web front-end (see~\cref{sec:frontend_restapi}). Based on available data (see~\cref{sec:datasources,sec:dataint}), simulations are performed, if possible using high-performace computing resources for fast and efficient computation (see~\cref{sec:automated,sec:hpc}).}
    \label{fig:modeling}
\end{figure}

In order to calibrate models on a local, district-scale, we run Bayesian inference using the probabilistic programming framework PyMC~\cite{AbrilPla2023PyMC}.
Similar to~\cite{Dehning2023}, we use a student's \textit{t}-distribution to define a robust likelihood of the observed case data that is robust to outliers and we incorporate regular change points for the contact rates using gradient-free DE Metropolis-Z~\cite{terBraakVrugt2008DEMCsnooker} sampling algorithm.
Convergence is checked using the rank-normalized $\hat{r}$ values~\cite{convergence_diagnostics}.

\subsubsection{Optimal control}
\label{sec:optimalcontrol}

Besides the simulation of scenarios, the MEmilio framework allows to leverage optimal control~\cite{betts2010practical} for ODE models to compute optimal intervention schemes, e.g., schedules of NPIs with optimized strengths, or optimized testing or vaccination strategies. 

While MEmilio supports optimal control for graph-based metapopulation models, the optimal control problem considered in our automated pipeline (see Section \ref{sec:automated}) models Germany without spatial stratification to reduce practical runtime. It can be written in its continuous-time formulation as
\begin{align}
\begin{aligned}
    \min_{\psi_l(t)} \quad & \int_{t_0}^{t_{\max}} \sum_{l \in \mathcal{L}} \psi_l(t) \ dt \\
    \text{subject to} \quad & \text{the }\textit{SECIRVVS}\text{ ODE model},\\
     & I(t) \leq I_{max}, \\
     & \psi_l(t) \in [0,1] \quad \forall l \in \mathcal{L},
\end{aligned}
\end{align}
where $\mathcal{L}$ is the set of considered NPIs, $\psi_l$ denotes the control variable of the $l$-th NPI, $I_{max}$ is the number of admissible infections at any point in time, and the objective function minimizes the cumulative exposure to NPIs over the optimization horizon $[t_0, t_{\max}]$. 
The control variable $\psi_l(t)$ modulates the effect of the $l$-th NPI on the contact patterns $\Phi(t)$, with $\psi_l(t)=1$ denoting full implementation and $\psi_l(t)=0$ no intervention effect (with linear scaling in between).

For implementation, the controls $\psi_l(t)$ are discretized as piece-wise constant segments with length of one week, thus allowing for weekly adaptation of NPI strengths, whereas the constraint $I(t) \leq I_{max}$ is evaluated on a daily basis. 
The optimal control problem is solved with IPOPT \cite{wachter2006implementation} using a direct single shooting approach, see, e.g., \cite{sass2025obscured}, and automatic differentiation \cite{naumann2011art}.

\subsection{Data integration}\label{sec:dataint}

\subsubsection{Requirements}

To integrate epidemiological datasets, from public repository and in-house databases, into a modeling framework, several requirements were identified. 
\begin{itemize}
    \item the data ingestion step should run periodically or on-demand by authorized external services,
    \item data must be stored according to data engineering standards to ensure efficient and reliable retrieval,
    \item a secure storage and access to private datasets must be implemented,
    \item the data integration interface must be scalable to enable efficient interaction with large datasets,
    \item communication and interoperability with other services should be ensured.
\end{itemize}

\subsubsection{General overview of MemiliSQL REST API}\label{sec:dataint-restapi}

The SARS-CoV-2 epidemiological datasets have been continuously gathered by German LHAs and shared with the Robert Koch-Institute (RKI) throughout the the pandemic. The data shared with the RKI was subsequently made available via the RKI public data repositories. To provide the needed datasets (see \cref{sec:datasources}) on demand to the modeling engine behind the platform, an Extract-Transform-Load (ETL) data pipeline was implemented. The pipeline periodically ingests and transforms data from hospitalization and intensive care units (ICU)~\cite{robert_koch_institut_2025_17623109}, vaccination~\cite{robert_koch_institut_2024_12697471}, and infection cases~\cite{robert_koch_institut_2025_17627234} from the RKI repositories using the MEmilio epidata package~\cite{Bicker_MEmilio_v2_1_0} before persisting raw and processed data in a managed datastore. In addition, demographic data from the Federal Agency of Work and the Federal Statistical Office of Germany (DESTATIS) is harvested~\cite{statistisches_bundesamt_destatis_population_nodate} and stored in the managed datastore. 

In addition to data from public repositories, richer data in LHA in-house SurvNet databases~\cite{survnet_db_rki} can be ingested and integrated in our solution. These datasets can contain features not present in public data. With the support of a pilot LHA, we 
created an SQL query to extract a set of epidemiological information, e.g., vaccination status, intensive care or general patient information. For data protection reasons, only authenticated LHA employees can run the query and upload the aggregated output CSV file using the upload functionality implemented in ESID; see~\cref{fig:esid-architecture}.  The uploaded data can then be sent to a staging location in MinIO File Storage awaiting transformation. To validate this feature, we first used real data provided by the pilot LHAs from times of high
pandemic activity. Second, we generated artificial individual pandemic data to prove the functionality of the whole
workflow. To generate individual data, we use MEmilio's agent-based model~\cite{KERKMANN2025110269}. Various features are used to assess data quality, perform data transformation, upload and extract the datasets into the database. Data security is ensured through our user management system, cf.~\cref{sec:usermanage}.

\begin{figure}
    \centering
    \includegraphics[width=1.0\linewidth]{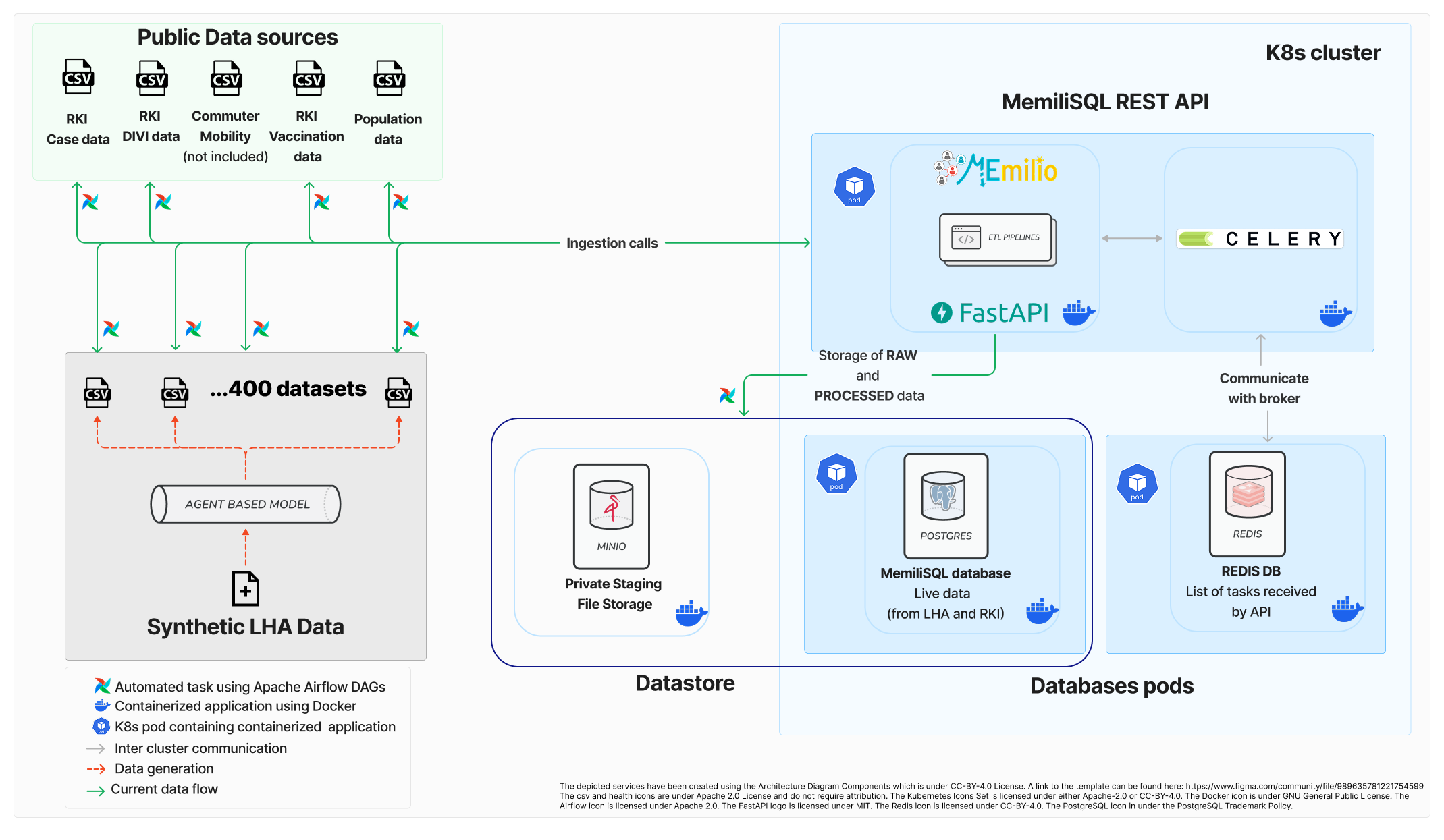}
    \caption{
    \textbf{A general overview of the data integration services.} The data integration services comprise a datastore and a REST API. Upon receiving requests, the API will proceed with the scheduling of various tasks, including the ingestion of specific dataset, its transformation and finally its upload into the MemiliSQL database.  
    }
    \label{fig:general-overview-memilisql}
\end{figure}

The ETL data pipeline was implemented as a standalone REST API built on FastAPI~\cite{fastapi}, coined MemiliSQL, and enhanced with Celery~\cite{celery} and Redis broker~\cite{redis} as task queue system. The MemiliSQL REST API allows other applications to trigger data pipeline steps. Each ETL data pipeline step, i.e., data ingestion, data transformation, data upload to the MemiliSQL database, and data extraction from the MemiliSQL database, has been defined as an API endpoint and can be requested using the \textit{GET} operation. Moreover, since data extraction and transformation steps are demanding in terms of computing resources and could not be handled by the MemiliSQL REST API, these were implemented as Celery tasks. These tasks can be scheduled to run asynchronously across different Celery workers within the Kubernetes pods dedicated to handle this workload, and their outputs can be temporarily stored in the Redis database. Offloading the REST API with a task queue allowed the API to remain responsive to incoming tasks while also scheduling their runs by workers. To create a fully automated and orchestrated data flow, we have implemented Airflow directed acyclic graphs (DAGs). Each DAG sends one or multiple API requests against the MemiliSQL REST API to create and initiate Celery tasks. The \cref{fig:general-overview-memilisql} gives a high-level overview of the ETL services that support the integration of both public and synthetic LHA datasets. A description of the different Airflow DAGs is given in~\cref{sec:automated}.   

\subsubsection{Deployment of the data integration services}\label{sec:dataint-deploy}
The architecture and infrastructure hosting the data pipeline is rather minimal. It comprises the MemiliSQL REST API and a datastore, which consists of a PostgreSQL database and MinIO File Storage. The deployment of applications (e.g., MemiliSQL REST API and database) is configured and managed by creating Fleet\footnote{\textbf{Rancher Fleet}. A container management and deployment engine that manage GitOps for a single Kubernetes cluster. \url{https://fleet.rancher.io/}} bundles from Helm charts\footnote{\textbf{Helm}. A Kubernetes-native package manager for templated and version-controlled application deployments. \url{https://helm.sh/docs/}}, which are stored in a Git repository. While the community PostgreSQL Helm chart was used for the MemiliSQL database, a tailored Helm chart  based on the containerized MemiliSQL REST API was created. Rancher Fleet is subsequently used to automate the deployment of these Fleet bundles as applications on a Kubernetes cluster. In addition to software application bundles, Fleet bundles were defined to create storage applications which provision and attach persistent volumes to Kubernetes database-containing pods. The provisioning of storage requires access to a computing infrastructure, and we used Openstack\footnote{\textbf{OpenStack}.
Open-source cloud computing platform for building private and public clouds. \url{https://www.openstack.org}}, an Infrastructure-as-a-Service (IaaS) platform, to host the Kubernetes cluster and provision CPU and RAM resources. Fleet continuously synchronizes the Kubernetes resources with the Git repository containing the Fleet bundles. This ensures the reproducibility and versioning of the overall infrastructure. Moreover, the deployment into a Kubernetes cluster enables reliable orchestration, provisioning, and management of computational resources. 

Data ingested from public data sources and data uploaded by LHAs are stored in the MemiliSQL database, which is a PostgreSQL database. The amount of storage needed for one day of data from public repositories varies. For one year of data collection, the storage needs to store raw and transformed datasets is estimated to be 500~GB.

\subsection{Front-End and REST API}\label{sec:frontend_restapi}

\begin{figure}
    \centering
    \includegraphics[width=0.8\linewidth]{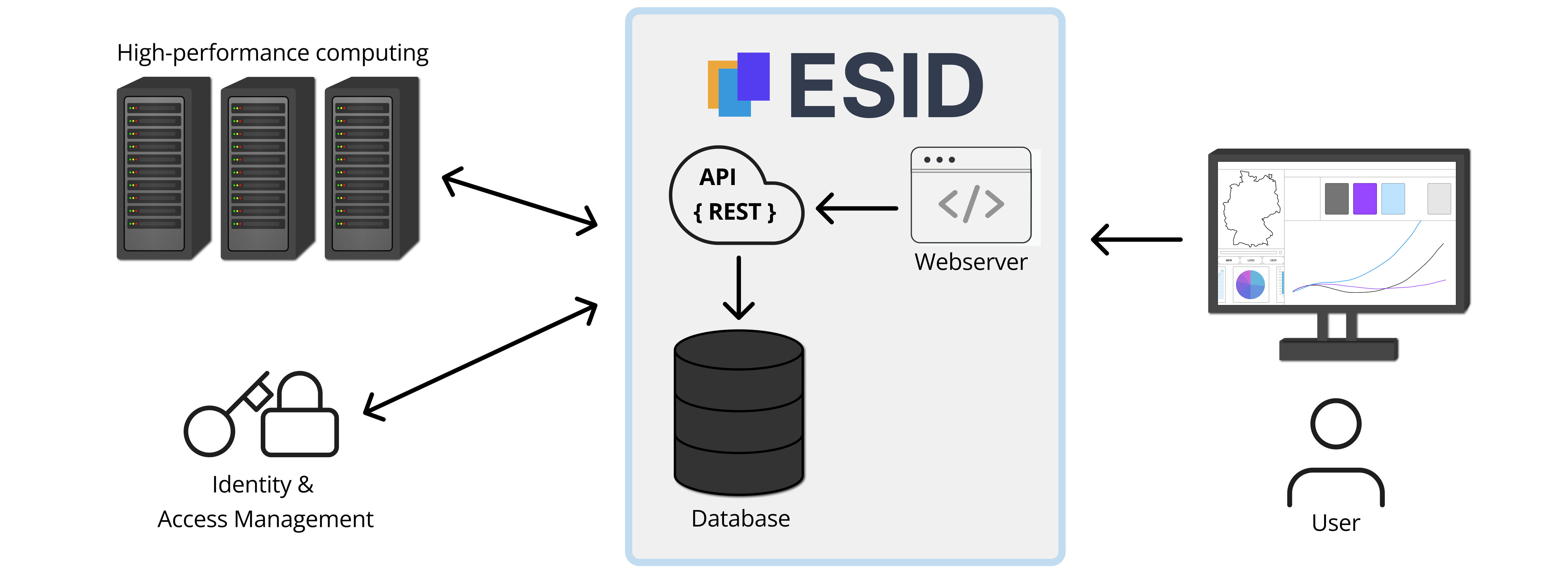}
    \caption{
    \textbf{ESID's architecture.} ESID's front-end can be accessed by the user and the ESID's back-end is accessed through the user query or an automated instance. The API verifies all actions with the identity and access management system.}
    \label{fig:esid-architecture}
\end{figure}

To ensure low-barriers for system access as well as swift and smooth adoption in pandemic situations, the web front-end ESID\footnote{\textbf{ESID Front-end}. The React-based website for ESID. \url{https://github.com/DLR-SC/ESID}} (\emph{Epidemiological Scenarios for Infectious Diseases}) was developed with a corresponding REST API in the background. \cref{fig:esid-architecture} shows the architecture of ESID and the other services it communicates with. The development was closely aligned with the guidelines reported in~\cite{betz2023}, in particular, addressing multiple user groups such as public health experts, decision makers, and individuals from the general public. By providing access to the general public, the platform fosters transparency and, thus, should increase acceptance for implemented interventions.

\subsubsection{Requirements}

The development of the user front-end was driven by the idea of a minimal barrier toolkit for operating and interconnecting with epidemiological models. This means that it should not require particular privileges to be installed or operated, and interfaces should be built upon standardized approaches. In line with current development in the public health sector in Germany~\cite{bmg_oegd}, we therefore opted for a web-based application. Many requirements for a visual analytics toolkit were already discussed in a recent publication~\cite{betz2023}, and a crucial additional step was the inclusion of four German health authorities that served as pilot LHAs to ensure that the final product was developed according to the users' needs.

We collected more than 100 user stories. From these, a handful of relevant requirements defined the choice of technology and architecture:
\begin{itemize}
    \item Retrospective and ongoing pandemic data need to be visualized,
    \item filtering in all dimensions of the dataset should be possible,
    \item interface response should be immediate, if possible,
    \item multiple user groups with different rights should be implemented,
    \item selected users should be able to trigger simulations with custom parameters,
    \item selected users should be able to upload data to be included in simulations.
\end{itemize}

\subsubsection{REST API}\label{sec:esid-back-end}

The ESID API\footnote{\textbf{ESID Back-end}. The back-end for the ESID website based on FastAPI and PostgreSQL. \url{https://github.com/SciCompMod/ESID-back-end}} serves as the critical middleware layer connecting the simulation back-end with the web-based front-end, facilitating seamless data flow and management within the system.

Built on FastAPI~\cite{fastapi}, the API provides a robust RESTful interface coupled with a PostgreSQL~\cite{postgresql} database for persistent storage and efficient data retrieval. Additionally, the API uses Celery web workers~\cite{celery} for more complex tasks, and stores temporary task results in an in-memory database using Redis~\cite{redis}. To optimize query performance, we employ strategic indexing of frequently accessed fields, ensuring rapid data retrieval even as the volume of simulation results grows. The API exposes two distinct categories of endpoints, each serving different architectural needs. back-end-facing endpoints handle model definition and data ingestion, facilitating integration with MEmilio (see~\cref{sec:modeling}) and Apache Airflow used for workflow orchestration (see~\cref{sec:automated}). These endpoints enable automated workflows to receive simulation requests, run simulations, and store results. Front-end-facing endpoints, conversely, provide optimized data access for visualization purposes, returning appropriately aggregated and filtered datasets based on user interaction, and submitting new simulations for approved users. The following endpoints exist to manage the complex and dynamic requirements:
\begin{itemize}
    \item \textbf{Models} are the foundation of all simulations. They define all model parameters; in which groups to split the population, and in which compartments the infection can be split into.
    \item \textbf{ParameterDefinitions} define the rules of the model, e.g., how likely it is for a person to become infected, how long the infection lasts, or how severe infections can get.
    \item \textbf{Groups} represent a flexible concept to stratify by age, gender, or socio-economic properties.
    \begin{itemize}
        \item \textbf{Categories} specify which groups belong together, e.g., the age category tracks all groups which specify an age range. Importantly, each person belongs to exactly one group of a category.
    \end{itemize}
    \item \textbf{Compartments} represent all the infection states of a model.
    \item \textbf{Nodes} are a generic way to define a spatial subdivision, generally given by administrative units.
    \begin{itemize}
        \item \textbf{NodeLists} group nodes together into a named set.
    \end{itemize}
    \item \textbf{Interventions} define disease mitigation measures each defined through a strictness and application area.
    \item \textbf{Scenarios} are the core of the application and combine the information above. Each scenario represents a simulated prediction, where a model with parameters, a set of nodes, a time frame, and interventions are chosen.
    \begin{itemize}
        \item \textbf{InfectionData} is the endpoint to get and filter the multidimensional data of a scenario. It offers parameters to filter by nodes, dates, compartments, groups, and percentiles.
    \end{itemize}
\end{itemize}

For all endpoints, at least \emph{GET}, \emph{POST}, and \emph{DELETE} operations exist.
The \emph{Scenarios} endpoint has an additional \emph{PUT} operation to upload simulation data.

The authentication system (see Section~\ref{sec:usermanage}) regulates which endpoints can be accessed with which permissions.
Generally, all \emph{GET} operations on endpoints are accessible without authentication; although they might only return results tagged for public access.
So, even though they are publicly accessible, they return different results depending on who requests the data.
\emph{POST}, \emph{PUT}, and \emph{DELETE} operations on endpoints always require authentication. \emph{Models}, \emph{ParameterDefinitions}, \emph{Groups}, \emph{Compartments}, \emph{Nodes}, and \emph{Interventions} can only be managed by system administrators or other authenticated services such as Apache Airflow (see~\cref{sec:automated}). Only authenticated users can create \emph{scenarios} and can subsequently share access with other users.

\subsubsection{Front-End}\label{sec:esid-frontend}

\begin{figure}
    \centering
    \includegraphics[width=0.9\linewidth]{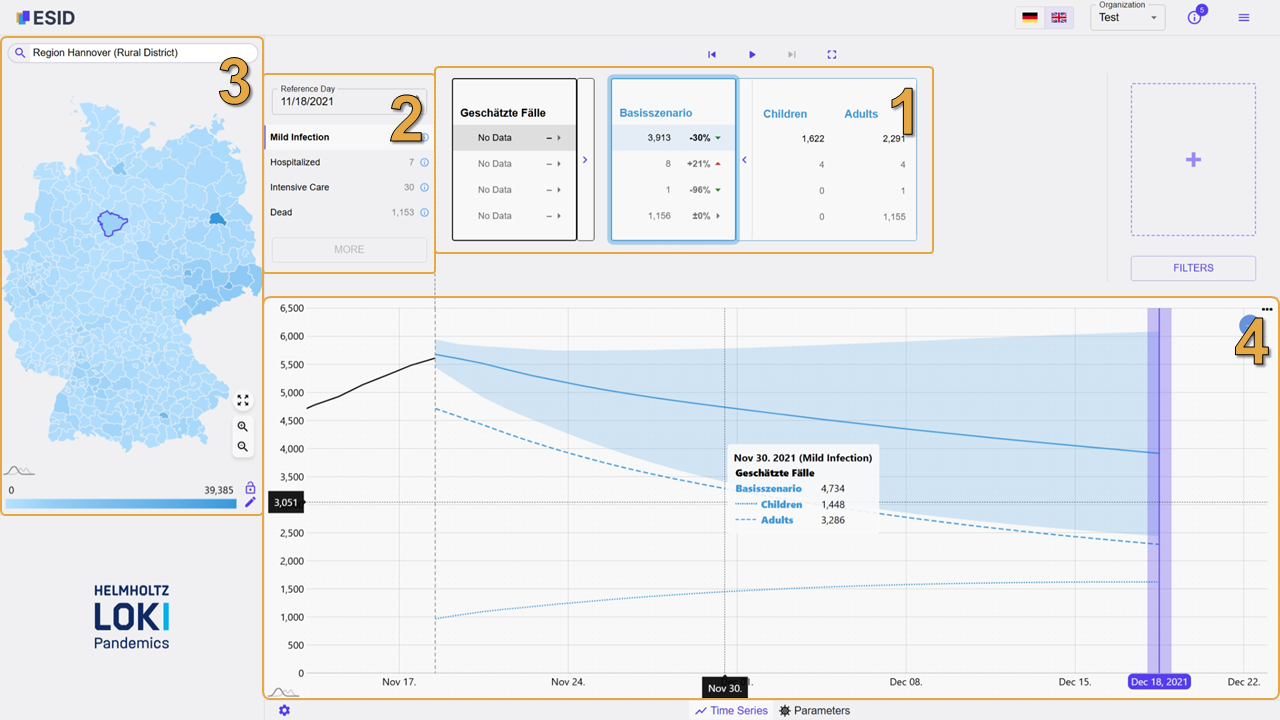}
    \caption{
        \textbf{Main user interface of ESID.}
        The numbered components each show a slice of the data cube, as well as serving as filters for the other components.
        1) The scenario cards show each scenario with data for the selected date, infection state, and node, including a percentage of change that shows the delta compared to the reference date; each can also be expanded for group specific data.
        Selecting a scenario changes the visualization of other components to show this scenario's data.
        2) A list of infection states (potentially grouped from compartments) that show data for the reference day displayed above the list. Selecting an infection state changes the other components to show data for this infection state.
        3) A map with regions (nodes) showing the spatial patterns of the selected data. Selecting a region changes the other components to show data for this region.
        4) A time series view showing the selected data.
        Contrary to other components, it can show multiple scenarios at once, to allow easy comparisons. For the selected scenario, additional information is shown with uncertainty bands (e.g., 25th and 75th percentiles) and dashed lines for group specific data.
        The vertical dashed line represents the reference day and the purple line the selected date. Selecting a day changes the other components to show data for that day. The small gap between the selected scenario (blue line) and the retrospective data (black line) can be explained by the comparably low number of runs conducted here and the stochasticity in the model initialization.
    }
    \label{fig:esid-ui}
\end{figure}

To fulfill the requirements of a modern and easy to use interface, the front-end was designed as a sophisticated visual filtering system based on brushing and linking principles~\cite{Keim2002}, enabling intuitive exploration of multidimensional simulation data. ESID is built on top of established frameworks, like React~\cite{react} and Redux~\cite{redux}, which allows the application to be highly interactive. The architecture is based on VVAFER, proposed by the authors of~\cite{zeumer2023}. The MUI library~\cite{mui} for interface components provides a modern look following the Material Design guidelines~\cite{material-design}. \cref{fig:esid-ui} shows the main interface of ESID, which contains four major connected components. Each component gives insight into at least one dimension (for example, time, space, and infection states) and actions on one component interactively as a filter and changes the view of the other components.

\begin{figure}
    \centering
    \begin{subfigure}{.71\textwidth}
        \centering
        \includegraphics[width=0.99\linewidth]{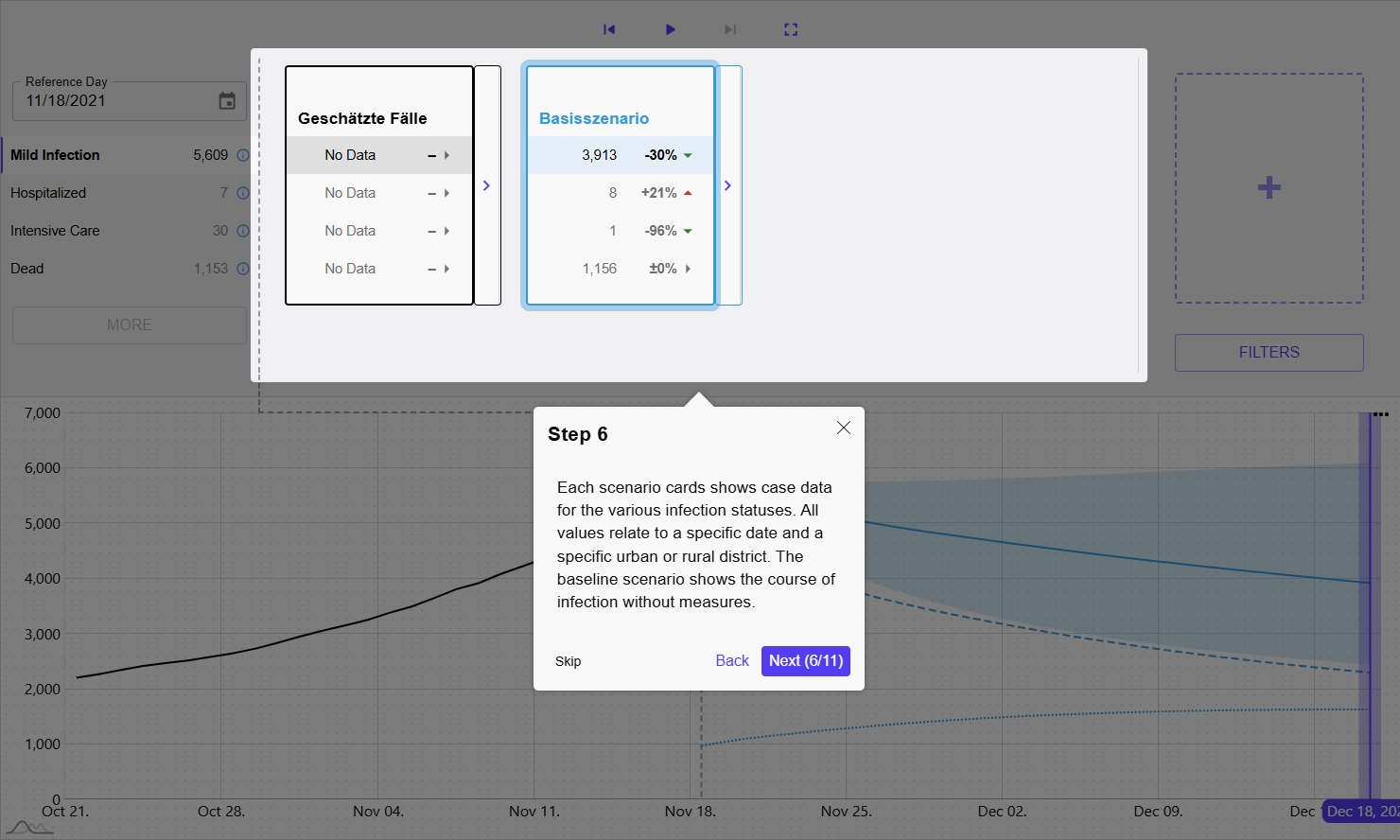}
    \end{subfigure}
    \begin{subfigure}{.277\textwidth}
        \centering
        \includegraphics[width=0.99\linewidth]{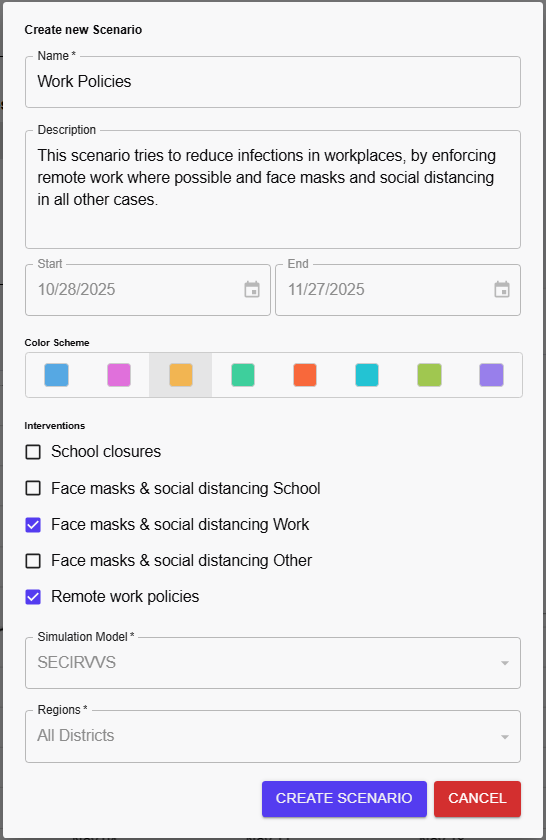}
    \end{subfigure}
    \caption{\textbf{Left: Excerpt of the interactive on-boarding of the front-end.} The scenario card view is highlighted and explained; other components are grayed out. \textbf{Right: Dialog to create a new scenario.}. The user can configure information for a custom scenario.
    }
    \label{fig:esid-onboarding-create-scenario}
\end{figure}

To make ESID as accessible and user-friendly as possible, several features have been added.
On first visit, an interactive on-boarding tour guides users through the platform's capabilities. By highlighting important aspects and encouraging the user to interact with the components, users can quickly familiarize themselves with each component's functionality; see~\cref{fig:esid-onboarding-create-scenario} (left).
The interactive tour can always be found in the help center which also provides a natural language based semantic search of related resources to answer users' questions.

A key technical challenge addressed in the front-end is state synchronization between the client application and the API. All selections by the user in the interface can be interpreted as a large filter for a data cube. The front-end maintains a local state that accurately reflects available data and current filters, and uses it to fetch the requested data efficiently. Data communication with the back-end is managed through RTK Query~\cite{rtk-query}, which implements intelligent caching strategies and pre-fetching to minimize latency and enhance user experience. These strategies store frequently accessed datasets locally, reducing redundant API calls and enabling instantaneous filtering operations on cached data.

A dedicated scenario creation interface allows authorized users to configure and submit simulations with custom parameters; see~\cref{fig:esid-onboarding-create-scenario} (right).
Every creation induces an interaction loop: A newly created scenario gets posted to the REST API. Airflow (see~\cref{sec:automated}) then submits all new scenarios to MEmilio (see~\cref{sec:modeling}) for simulation and writes the results to the REST API.
After simulation of the scenario, the user can evaluate the effects of the contact reduction through the NPIs, drawing individual conclusions and submit further scenarios. Authentication flows in the front-end mirror the API's tiered approach, seamlessly integrating with the identity provider (IDP) presented in~\cref{sec:usermanage}.

\subsection{Authentication and authorization system}\label{sec:usermanage}
The public front-end from the previous section allows our users to trigger, visualize, and analyze simulations of infectious disease spread. These simulations might need substantial computational resources and may contain sensitive data that was uploaded by an LHA but is publicly undisclosed. It is therefore crucial that only authorized entities can access specific functionalities. We implement an authentication and authorization system that allows us to cater to a wide range of potential user groups, e.g., the general public, LHA employees, and public health experts.

\subsubsection{Requirements}
\begin{wrapfigure}{R}{0.4\textwidth}
    \opensans
    \centering
    \resizebox{0.4\textwidth}{!}{
        \begin{tikzpicture}
    
            \node[align=center] (Loki) at (7, 12) {{\Huge \faUser} \\ \textbf{\small Project Staff}};
    
            \node[align=center] (IT1) at (7, 9) {{\Huge \faUser} {\Large \faDesktop} \\ \textbf{\small LHA IT Staff}};
    
            \node[align=center] (IT2) at (12, 9) {{\Huge \faUser} {\Large \faDesktop} \\ \textbf{\small LHA IT Staff}};
    
            \node[align=center] (MED1) at (7, 6) {{\Huge \faUserMd} \\ \textbf{\small Public health Staff}};
    
            \node[align=center] (MED2) at (12, 6) {{\Huge \faUserMd} \\ \textbf{\small Public health Staff}};
    
            \draw[fill=lightgray, opacity=0.3, dashed] (9, 5) -- (5, 5) -- (5, 10) -- (9, 10);
            \node[align=center] (LHA1) at (7, 4.5) {\textcolor{darkgray}{\textbf{LHA No. 1}}};
    
            \draw[fill=lightgray, opacity=0.3, dashed] (10, 5) -- (14, 5) -- (14, 10) -- (10, 10);
            \node[align=center] (LHA2) at (12, 4.5) {\textcolor{darkgray}{\textbf{LHA No. 2}}};
    
            \draw[->, line width=1.5pt] (Loki) -- (IT1);
            \draw[->, line width=1.5pt] (Loki) to [out=0, in= 90] (IT2);
            \draw[->, line width=1.5pt] (IT1) -- (MED1);
            \draw[->, line width=1.5pt] (IT2) -- (MED2);
    
        \end{tikzpicture}
    }
    \caption{
        \textbf{The different users of ESID and their relationships.} The project staff administers local IT staff who, in turn, administer their public health experts.
    }
    \label{fig:user-relations}
\end{wrapfigure}
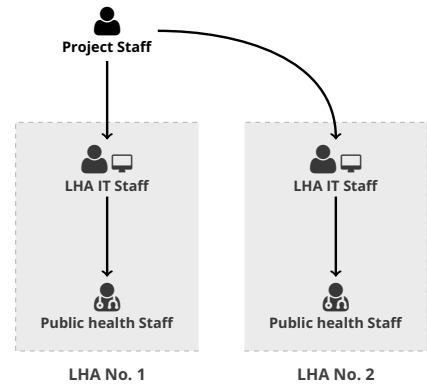

We first identified the need for different user roles and categorized the system's users into three user groups: the general public, IT staff from LHAs, and public health experts from LHAs. The latter then also serve as connection point to decision makers. For the general public no requirements apply and they can freely use the web front-end as a source of information. To manage the user base, the IT staff of LHAs is allowed to administer the local health experts; we summarize the relations between user types in \cref{fig:user-relations}.

We furthermore require a system that allows for the decentralized on-boarding and management of users at different LHAs, fine-grained access control of our sensitive API, and resource sharing as well as collaboration between different LHAs and expert users. Eventually, the system should be open to extension and integration with other identity providers such as the German BundID~\cite{bundid}. To generally address the requirement of a secure authorization and authentication system, standardized and state-of-the-art protocols should be used.

\subsubsection{Implementation}
To implement our authentication and authorization system, we use the OAuth~2.0
authorization framework~\cite{oauth2} and the OpenID~Connect authentication
protocol~\cite{OpenIDconnect} together with a self-hosted Keycloak~\cite{keycloak} server as our identity provider. 

\paragraph{OAuth~2.0}
OAuth~2.0 is an authorization framework that allows a third-party application to obtain limited access to a (HTTP) service on behalf of a consenting user without exposing the user's credentials or identity to the application. For instance, a user can grant a printing service access to their personal cloud data, without sharing their credentials with the printing service.

Formally, OAuth~2.0 defines four roles: resource owner, client, authorization server, and resource server. The resource owner is an entity capable of granting access to a protected resource. Usually, this is a person and we refer to them as end-user. The client is an application that makes requests to access protected resources on behalf of the resource owner with their authorization. In practice, the client is often a web-application running in the browser and the end-user authorizes the client by logging into the authorization server. The authorization server issues access tokens to the client after it successfully authenticated the resource owner and obtained authorization. Lastly, the resource server is hosting the protected resources and responds to requests to the protected resources using access tokens. In our use case, the resource owners are the public health experts at LHAs, the client is the web front-end, the authorization server is our self-hosted Keycloak instance, and the resource server is the back-end REST API.

\begin{figure}
    \fontfamily{qcr}\selectfont
    \centering
    \resizebox{\textwidth}{!}{
    \includegraphics[width=0.35\linewidth]{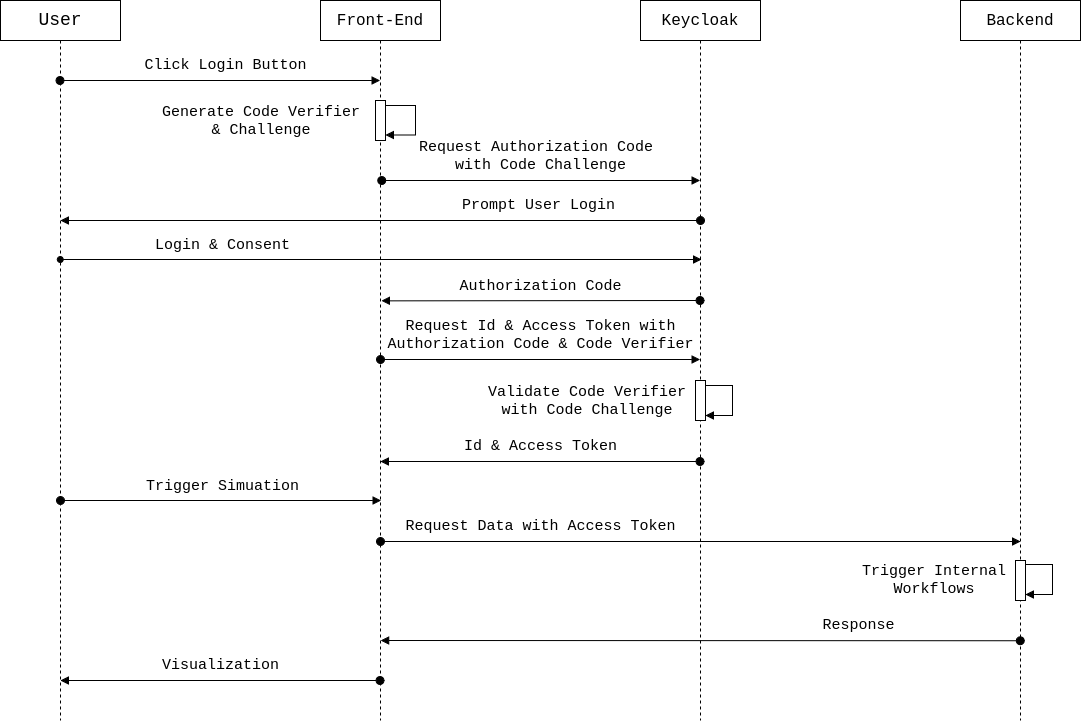}}
    \caption{
        \textbf{Overview of the login flow of ESID.} First, a user logs into ESID by clicking the login button, which starts the PKCE flow between ESID and our Keycloak instance. When ESID exchanges the authorization code and code verifier for the ID token and access token, the flow is finished and ESID
        can now make requests to the back-end API on behalf of the user potentially triggering internal workflows as described in \cref{sec:automated}.
    }
    \label{fig:esid-login-flow}
\end{figure}

To increase its flexibility, OAuth~2.0 provides so-called flows, each specifying an authorization protocol. Since an in-depth discussion of each flow is out of scope for this work, we refer the reader to its specification~\cite{oauth2} for more details, and, instead, focus our attention on the flow we chose for ESID: the Proof Key for Code Exchange (PKCE) flow. Defined in~\cite{pkce}, the PKCE flow is an extension of the authorization code flow that protects clients against the interception and the injection of authorization codes by introducing a random code challenge in each request. We chose this flow as it is the recommended flow with the best security guarantees for single page applications like ESID. \cref{fig:esid-login-flow} shows the login flow of ESID.

To explain the PKCE flow, we will now focus on the communication between the web front-end (the client) and our Keycloak instance (the authorization server).
The goal of the flow is to let the web front-end act on behalf of the end-user using it such that it can, for instance, make calls to APIs that are only available to public health experts. To achieve this, the web front-end will first create a code verifier by generating a random high-entropy string, and a code challenge by hashing the code verifier. Then, the web front-end will request an authorization code from Keycloak and also transmit the code challenge. If the end-user logs in, i.e., authorizes the web front-end, Keycloak will return an authorization code. The authorization code is a unique, short-lived, single-use token that Keycloak associates with the request, in particular, with the code challenge. Next, the web front-end will exchange the authorization code and code verifier for an access token, which it can then use to make requests on behalf of the end-user. To do so, the server will check that the code challenge it associated with the authorization code is really the hash of the code verifier and, if so, respond with an access token. Once the web front-end obtained the access token, it will send it along any requests to the web front-end API which will verify the token using the public key of the Keycloak realm the user is logged into, and, if successful, allow the web front-end to act on behalf of the end-user.

\paragraph{OpenID~Connect}
OpenID~Connect~\cite{OpenIDconnect} is an identity layer built on top of the OAuth~2.0 protocol that allows a client to verify the identity of an end-user established by authentication with an authorization server. We refrain from formally introducing and outlining how OpenID~Connect works but instead only describe how it builds upon the PKCE flow we have chosen.

To allow for authentication, OpenID~Connect uses an ID token which is issued by the authorization server and that contains information about the end-user and their authentication status with the authorization server, e.g., which realm they are logged into and which roles they have. In the case of the PKCE flow, this token is returned together with the access token when exchanging an authorization code. The token is then sent along with the access token to the REST API which can use it to, for instance, check whether the user is logged in, and to check which LHA the user belongs to.

\paragraph{Keycloak}
Keycloak is an open-source identity and access management software supporting standard protocols such as OAuth~2.0 and OpenID~Connect, two-factor authentication, as well as permission and role-based access management. In our system, Keycloak possesses the role of the identity provider (Idp) or, to put it in the terms of OAuth~2.0, the authorization server. That is, our users have an account at our Keycloak server, it stores their credentials, and, subsequently, authenticates them for our other services. We self-host the Idp is to ensure ownership of the user's data and to not rely on other Idps which might not operate under EU data protection laws.

To model the relationships between our users as depicted in \cref{fig:user-relations}, we use Keycloak's role-based access management and its concept of \emph{realms}. A realm is a logically isolated identity provider for a single application. That is, each realm holds its own user base, which cannot interact with the users of other realms. The main purpose of realms is to allow for a single Keycloak instance to serve multiple, independent applications. However, we use realms to isolate the users of different LHAs by creating one realm per LHA, achieving our goal of making it impossible for admins of one LHA to interact with the staff of other LHAs. To give admins the rights to manage the public health experts at their LHA, we use a simple role-based permission system which allows users that have the admin role. To protect our API, we can then check whether a user is logged in with our Keycloak server, which realm (i.e., LHA) they belong to, and whether they have the correct permissions.

\subsection{Pipelines and workflow orchestration}\label{sec:automated}

To provide decision makers and public health experts with accurate monitoring and time-critical decision support, up-to-date models (see~\cref{sec:modeling}), daily available data (see~\cref{sec:datasources,sec:dataint}), and the accessible front-end (see~\cref{sec:frontend_restapi}) need to be supplemented by automatically executed workflows that substantially reduce manual intervention loops. To this end, our platform adopts state-of-the-art workflow orchestration practices using Apache Airflow, ensuring that workflows of multiple steps are run in the right order, at the right time, and on the right hardware. Reliability and reproducibility at scale are achieved through a cloud-native deployment of Apache Airflow on a Kubernetes cluster managed via Rancher, running on an OpenStack-based IaaS layer. This setup provides modular integration of additional hardware and software components as needed. Furthermore, Unity-IDM\footnote{\textbf{Unity-IDM}. Identity and Access Management platform supporting federated authentication and fine-grained user management. \url{https://unity-idm.eu}} is used for federated, institution-backed authentication to provide controlled access and fine-grained user management for both development and operation. Finally, MinIO\footnote{\textbf{MinIO}. High-performance, open-source, S3-compatible object storage system. \url{https://www.min.io}} provide workflows with an S3-compatible persistent object storage back-end MinIO, enabling multiple workflows, even running on heterogeneous hardware, to efficiently share and reuse the same datasets. 

\subsubsection{Requirements} 

The implementation of our pipeline or workflow is driven by a set of requirements that ensure automation, scalability, robustness, and reliability in managing epidemic mitigation workflows. These requirements cover both functional and nonfunctional aspects, supporting management and processing of data in heterogeneous computational environments. In the following, we provide a high-level overview of the requirements.

\begin{itemize}
    \item Provide end-to-end automation from data collection to result dissemination, with reproducible, traceable execution without manual intervention,

    \item support dynamic scalability to accommodate varying computational loads and capability to allow new models, data sources, and sinks to be integrated.

    \item ensure reliable storage, sharing, and versioning of intermediate and output data in accordance with FAIR (Findable, Accessible, Interoperable, Reusable) principles~\cite{wilkinson2016fair},

    \item enforce fine-grained, federated authentication and authorization for shared services (e.g. pipeline source code, object storage), based on institute-backed identity assurance, with adequate access, modification, and deployment of workflow resources and stored results to authorized storage back-ends,

    \item provide continuous operation with acceptable performance, minimal downtime, and resilient to transient failures,

    \item interoperate with institutional and external infrastructure services, and compatible across containerized environments,

    \item capture and expose detailed workflow execution metadata to support runtime analysis and auditing.
\end{itemize}

\subsubsection{Architectural Realization}
\begin{figure}[h] 
    \centering
    \includegraphics[width=0.9\textwidth]{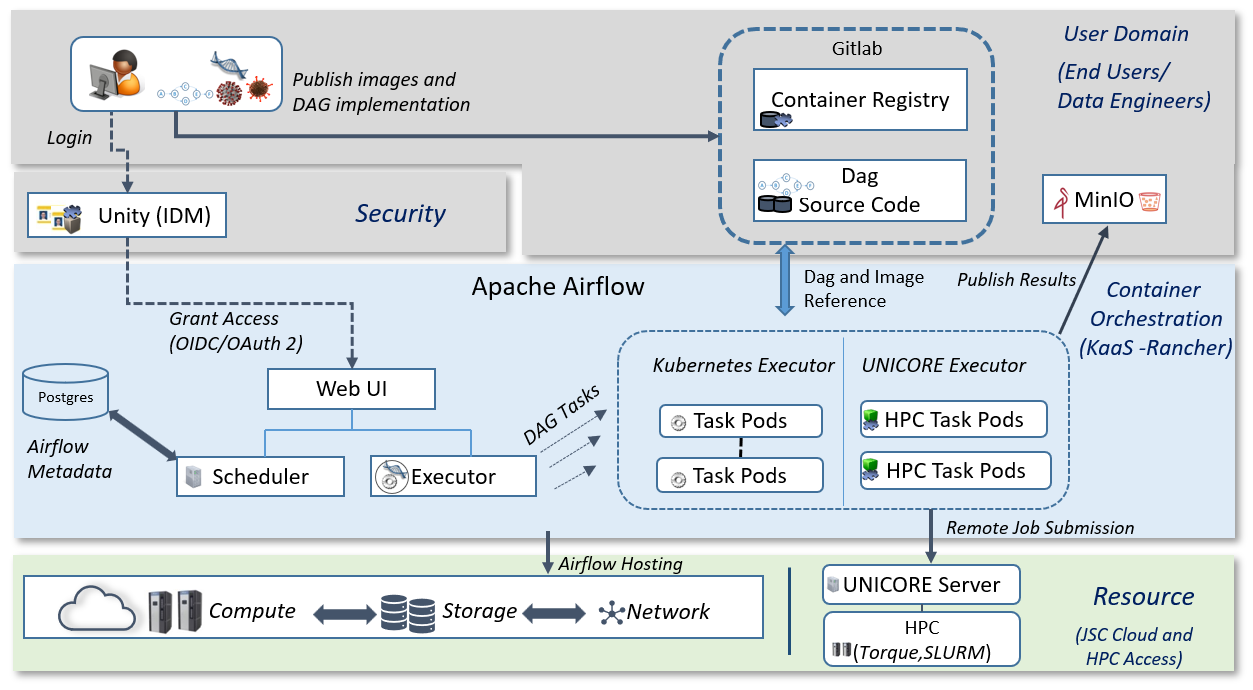}
    \caption{\textbf{Automated workflows architecture}. This figure shows the User domain (grey section), Container Orchestration tier (blue section) and Resources or IaaS tier (green section) which also highlights the integration with the HPC environment via UNICORE.}
    \label{fig:airflow-architecture}
\end{figure}

The architecture of the LOKI pipeline brings together workflow orchestration and enactment, container execution, data management, and authentication into a unified automation environment to support the visualization of epidemic developments. It is organized into three tiers -- the User Domain, the Container (Platform) Domain, and the IaaS Domain -- which separate workflow logic from orchestration and infrastructure management (see~\cref{fig:airflow-architecture}). This architecture improves maintainability, scalability, and security across distributed environments and extends the foundational concepts introduced by~\cite{Memon2024}.

At its core, Apache Airflow defines workflows as directed acyclic graphs (DAGs) where nodes represent individual tasks and edges define dependencies. The scheduler regularly evaluates all active DAGs, determines tasks whose dependencies have been satisfied, and submits them for execution. The executor provides the bridge between Airflow's orchestration logic and the underlying compute environment. In our deployment, we use the executor of type KubernetesExecutor, which is used to run tasks in a Kubernetes-based environment. This executor dynamically creates a dedicated \textit{Task Pod} for each task. A \textit{Task Pod} is an ephemeral instance in a containerized environment with its own resource limits (CPU and memory), mounted volumes, and isolated runtime context. Upon task completion, the corresponding \textit{Task Pod} is automatically terminated, allowing cluster resources to be released and reused efficiently. 

All workflow elements, including DAGs, executors, and configuration templates, are maintained in a central GitLab repository following a workflows-as-code paradigm. Airflow automatically synchronizes these elements via a Git-based deployment mechanism, which allows consistent versioning across development and production environments. The Airflow deployment configuration, such as container image references, runtime parameters, and synchronization policies, is managed declaratively using Helm charts\footnote{\textbf{Helm}. A Kubernetes-native package manager for templated and version-controlled application deployments. Documentation available at: \url{https://helm.sh/docs/}.} in accordance with Infrastructure-as-Code (IaC) principles~\cite{RAHMAN201965}. The Helm chart itself is version-controlled in GitLab. While keeping both the workflow code and the deployment configuration in Git, the pipeline and the Airflow deployment can be tracked and reproduced consistently across environments.

The User Domain (see \cref{fig:airflow-architecture}, grey section) is comprised of workflow authors, data scientists, and service accounts interacting with Airflow through its native web interface or REST API. Authentication and authorization are handled by Unity-IDM using OpenID~Connect and OAuth~2.0 standards, providing federated and interoperable access control consistent with institutional defined identity policies. Within this layer, the Airflow webserver was extended to support custom role handling and dynamic redirect URI processing during authentication.

The Container Orchestration tier (see \cref{fig:airflow-architecture}, blue section) hosts Airflow components—scheduler, webserver, and executor—alongside dynamically created taskpod instances. The underlying deployment is based on a Rancher-managed Kubernetes cluster, which handles orchestration, scaling, and lifecycle management. As mentioned above, each taskpod is assigned explicit CPU and memory requests and limits to achieve fair workload distribution. Intermediate workflow tasks communicate through the shared MinIO object store instance for data exchange and storage of intermediate or final results.

Workflow metadata including DAG definitions, task states, configurations, and execution logs are maintained within the same platform layer. This provenance metadata is stored in a centralized PostgreSQL database instance providing traceability and auditability of workflow executions. The execution logs are aggregated on a shared Network File System (NFS) volume mounted across Airflow components. Since logs from daily runs can grow rapidly, retention and cleanup policies are applied to prevent storage saturation and help maintaining stability across nodes.

The Resource tier (see \cref{fig:airflow-architecture}, green section), which provides the underlying infrastructure. This domain relies on an OpenStack-based IaaS, provisioning compute, storage, and networking resources for all Kubernetes nodes. Each Kubernetes node consumes a virtual machine configured via OpenStack flavors that define CPU, RAM, and ephemeral storage. The PostgreSQL database, object store, and NFS server are also hosted within this domain. Depending on the actual capacity of the infrastructure, resource quotas and volume allocations for the Airflow deployment are explicitly defined at this layer. 
This tier also highlights the connection to HPC resources through UNICORE~\cite{Unicore2005}, enabling Airflow to delegate compute-intensive simulation tasks to remote HPC resources whenever their runtime requirements exceed the limits of Kubernetes deployment. Details on the integration of external HPC resources via UNICORE (see~\cref{sec:hpc}).

\subsubsection{ESID pipeline}
\begin{figure}[h] 
    \centering
    \includegraphics[width=0.85\textwidth]{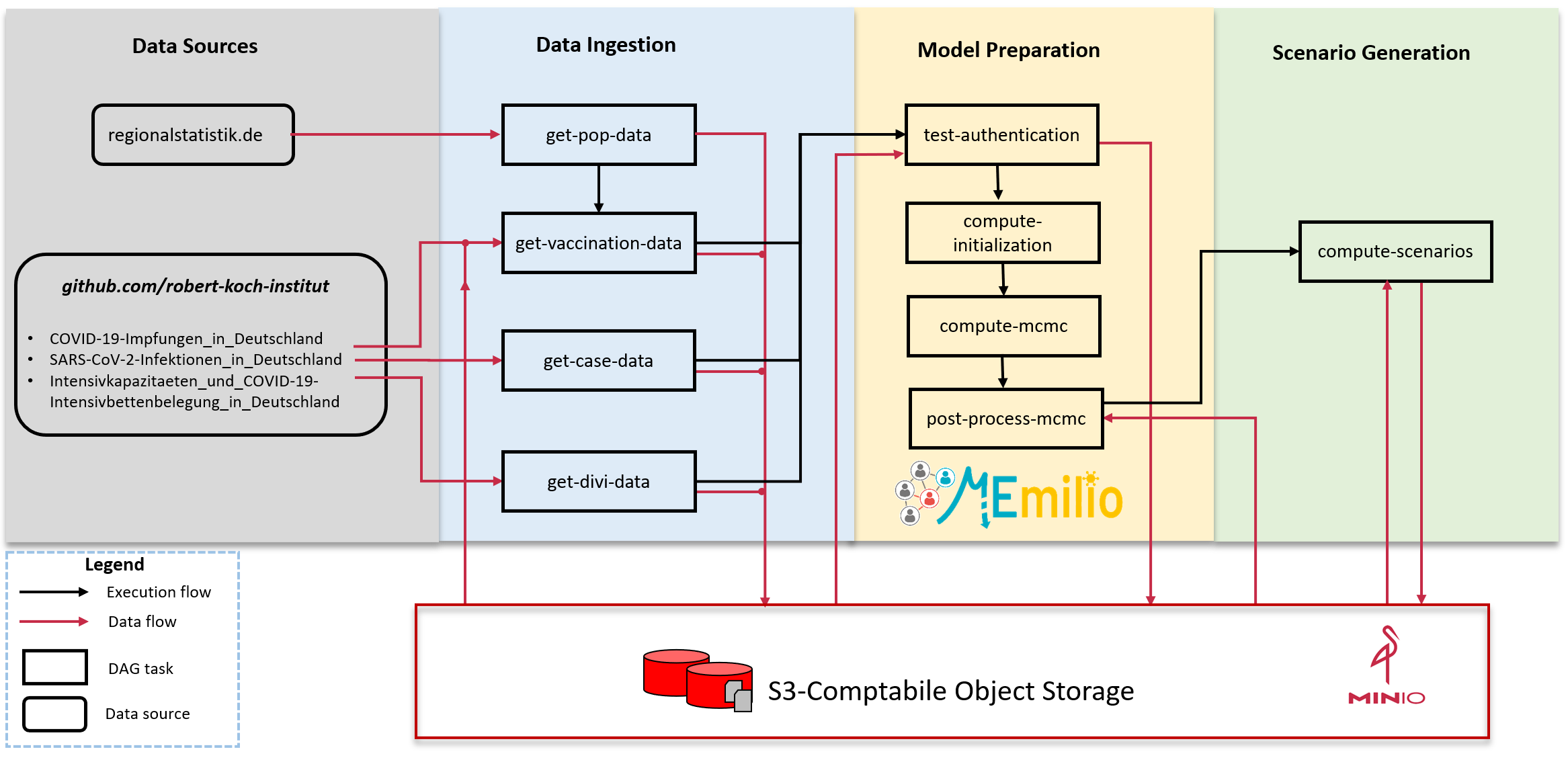}
    \caption{\textbf{ESID Pipeline consisting of three phases.} The figure shows the high-level workflow phases of data ingestion, model preparation and scenario generation including post processing and results upload, as implemented in Apache Airflow. The pipeline integrates data sources, distributed computation, and S3-compatible storage via MinIO.} 
    \label{fig:airflow-pipeline}
\end{figure}

The \textit{ESID pipeline} is a core workflow that automates data management and processing of reported and simulated data. It is composed of several data- and compute-oriented steps, including data collection, model preparation, and scenario-based forecasting. The data ingestion logic and underlying ETL processes are described in detail in \cref{sec:dataint-restapi}. \cref{fig:airflow-pipeline} provides a visual depiction of the ESID pipeline, which is structured into three main phases: data ingestion, model preparation, and post-processing with results upload. 

The \textit{Data Ingestion} phase consists of four tasks that retrieve heterogeneous epidemiological and demographic datasets from public sources. In practice, reported COVID-19 incidence and case statistics are obtained from the Robert Koch Institute (RKI) through the \texttt{get-case-data} task. The \texttt{get-divi-data} task retrieves information on ICU occupancy. Vaccination coverage is collected using \texttt{get-vaccination-data}. In addition, \texttt{get-pop-data} loads population and demographic distributions for German administrative regions. The resulting datasets are harmonized into a unified internal representation for downstream processing. 

The subsequent \textit{Model Preparation} phase begins with the test connection task (\texttt{test-authentication}), which verifies connectivity to remote compute and storage environments hosted at the Container Orchestration tier (depicted in \cref{fig:airflow-architecture}). The \texttt{compute-initialization} task aggregates all inputs and constructs the compartmental model structures required for parameter calibration. These structures are then used by the distributed \texttt{compute-mcmc} tasks, which perform one global parameter estimation task using Markov Chain Monte Carlo sampling. This phase concludes with the \texttt{postprocess-mcmc} task, which is responsible for generating posterior summaries and persisting intermediate results for subsequent analysis.

The \textit{Scenario Generation} phase represents the \texttt{compute-scenarios} task, which takes the posterior summaries from the previous phase and simulates epidemic trajectories under varying parametric assumptions. It finally uploads the resulting data to the S3-compatible storage system (MinIO) for visualization and decision support. Furthermore, the task includes an optimal control scenario configured as described in Section \ref{sec:optimalcontrol}, whose optimized control variables are subsequently applied to a simulation with district-level resolution as a basis for results visualization.

The ESID pipeline runs in an Airflow--Kubernetes environment and is subject to several practical constraints that affect scheduling and runtime behavior. Since the pipeline is executed as part of a daily reporting process, each run must finish within a fixed time window. This limits the amount of computation that can be performed per run, even when additional HPC resources are available. Because workflow tasks are executed as individual Kubernetes pods, the level of parallelism depends on the cluster's available CPU, memory, and node capacity. When many tasks run concurrently, this can lead to scheduling delays or pod evictions.

Some pipeline steps also depend on external data sources and institutional authentication services. Variations in availability or response times of these services can delay pipeline startup or execution. In addition, the pipeline is deployed in a distributed setup, where Airflow DAGs and container images are stored in a remote GitLab registry. Pulling updated images from external sources can introduce startup delays, especially when node caches are cold. Finally, repeated execution of containerized tasks can lead to the accumulation of cached image layers and other temporary data on compute nodes. If this data is not cleaned up, it can degrade performance over time.

\subsection{Automation combined with High-Performance Computing}\label{sec:hpc}

For computationally heavy and highly parallelizable workflows, access to distributed HPC resources is essential to produce near-time results, e.g., in daily analyses and short-term prediction of infection case numbers. Achieving this requires a middleware that connects Airflow to HPC systems. In the context of this work, the HPC system of our software stack should interact with JURECA~\cite{JURECA}, a pre-exascale modular supercomputer operated by the Forschungszentrum Jülich. JURECA offers several types of compute nodes. Daily analyses, inference, and short-term prediction are all tasks suitable for the standard compute nodes offered by JURECA. Each standard compute node offers two AMD EPYC 7742 processors (2×64 cores running at 2.25\,GHz) and 512\,GB of DDR4 memory.

\subsubsection{Requirements}\label{sec:hpcreqs}

For a middleware that connects HPC resources and automated workflow orchestration, we identified the following requirements:
\begin{itemize}
    \item open, well-documented standards-based APIs for remote access to HPC resources, 
    \item secure, auditable authentication and authorization,
    \item support of complete job and data lifecycle management, 
    \item stable service endpoints suitable for long-term operation,
    \item integration with institutional identity management systems and HPC schedulers,
    \item uniform access to heterogeneous HPC resources.
\end{itemize}

\subsubsection{HPC Integration}\label{sec:hpcimpl}
To satisfy the HPC integration requirements outlined above, UNICORE~\cite{Unicore2005, Unicore2016} is used as the middleware layer connecting Apache Airflow to external HPC resources. UNICORE is an open-source, state-of-the-art middleware stack that provides a uniform abstraction over heterogeneous HPC infrastructures and schedulers, enabling portable computation across a wide range of HPC systems. 

To enable Airflow to interact with HPC resources via UNICORE, a dedicated Airflow executor, the UNICOREExecutor, is provided 
by the \emph{airflow-unicore-integration} project~\cite{airflow-unicore-integration}. The project is released under a permissive BSD-3-clause license via PyPI. The UNICOREExecutor comes with a set of operators built around a common base, the \texttt{UnicoreGenericOperator}, which implements core functionality for job submission, monitoring, status retrieval, and output collection. Specialized operators extend this base class and provide convenience presets for different submission modes, such as staging and executing user scripts or submitting batch job scripts. In this work, batch jobs are submitted using the \texttt{UnicoreBSSOperator}, which extends \texttt{UnicoreGenericOperator} and stages a BSS script for execution via UNICORE.

By integrating the UNICOREExecutor, the ESID pipeline can offload computationally or memory intensive tasks, such as large-scale simulations, to HPC resources. This allows the pipeline to execute workloads exceeding the resource limits of a Kubernetes-based deployment.

\section{Results}\label{sec:results}

\subsection{A full software stack for epidemic disease management}\label{sec:results_framework}

\begin{figure}[!t]
    \centering
    \includegraphics[width=\textwidth]{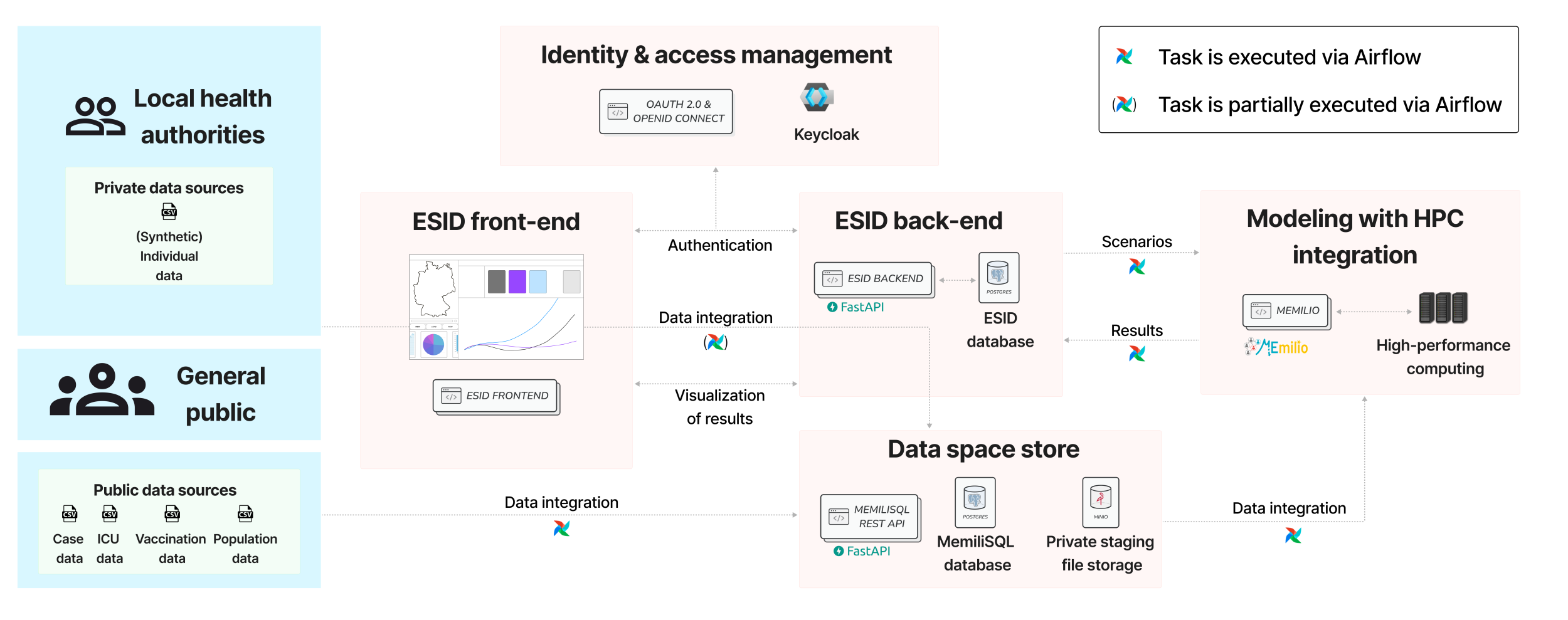}
    \caption{\textbf{Overview of full the software stack.} On the left, we visualize public and potential LHA-specific data sets as well as users from the LHAs and the general public. Upon interaction with the ESID front-end and authentication, private data sources can be integrated. Furthermore, any interaction with the front-end triggers actions in the ESID back-end to retrieve data from the data store or to send scenario requests to the modeling and simulation stage with HPC.}
    \label{fig:overview_software_stack}
\end{figure}

The core result of this paper is the full stack of scientific software as visualized on a high-level in~\cref{fig:overview_software_stack}. It enables users of the ESID web front-end to interact and filter reported case and simulated data and to allow for swift exploration of “what-if” questions. In regular intervals, public data as well as queued HPC simulations are fed to the database via the ETL pipeline. To include local data, LHAs can upload individual data sets via the front-end which is then transferred to the ESID back-end and integrated into the database. A proof-of-concept for the integration of private or local data is given in~\cref{sec:local_data_scenario}. Aside from a predefined set of scenarios that cover common intervention settings such as 50~\% remote work or mask wearing in public settings, individual scenarios from authenticated users can be stored in the ESID back-end. To ensure data privacy, only authenticated users can upload data and view scenarios that were computed using individual data. In addition to scenarios with common interventions, an optimal control scenario is preconfigured. By default, the model is initialized with public data as in the pilot phase with only four LHAs, individual data is i) not available for the large majority of districts and ii) for any individual contribution, it is necessary to define if the data can be shared with other LHAs or districts. Upon updates of the integrated data, the scenarios are computed through MEmilio's metapopulation model, using the HPC resources to allow for a fast completion of tasks. The results are passed back to the ESID back-end and eventually visualized in the front-end. Whenever possible, tasks are executed automatically via Airflow to keep interactive latency low while maintaining traceability of every published scenario (inputs, seeds, image tags). For the open-source publication of the individual software components, see the section on software and data availability.

\subsection{Proof-of-concept for LHA data integration}\label{sec:local_data_scenario}

\begin{figure}
    \centering
    \includegraphics[width=1.0\linewidth]{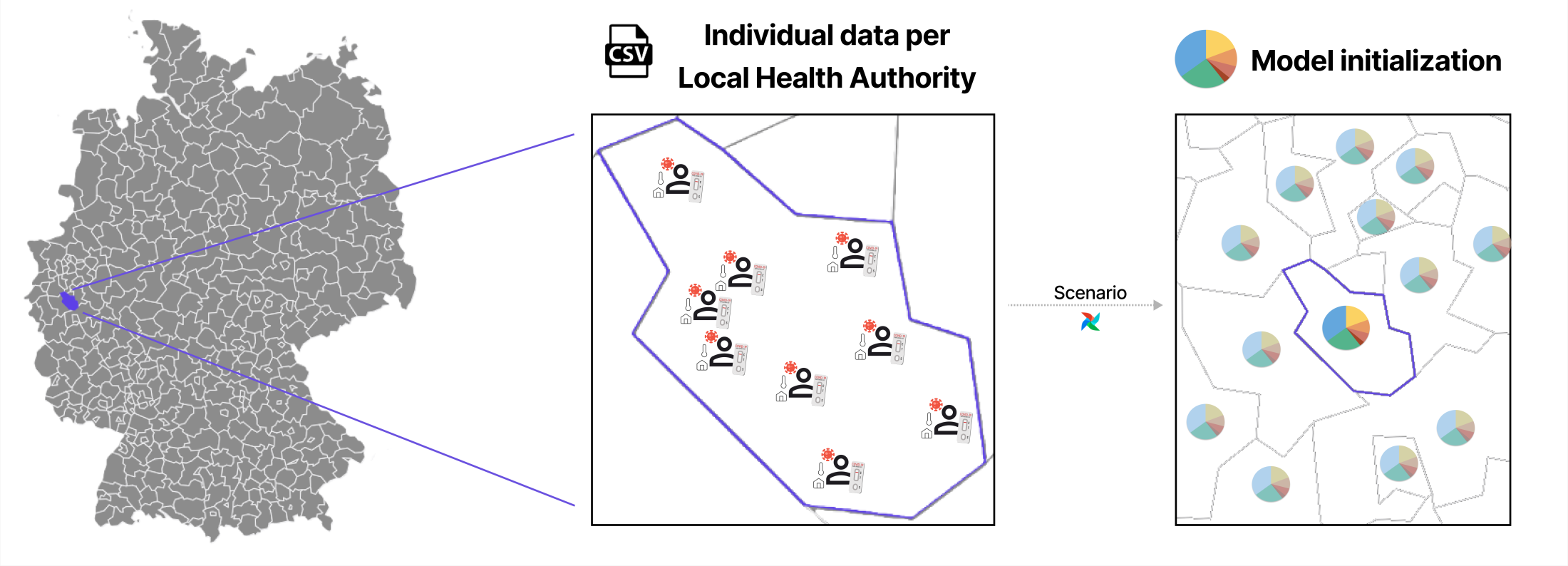}
    \caption{\textbf{LHA data integration and use in Germany.} Available LHA data on individual level is used to improve accuracy and granularity in scenario simulations. Availability for data in the exemplary district of Cologne is shown in the left part of the figure. The improved initialization of a metapopulation model in this district is depicted by the brighter display of the population cohorts in the selected district in the right picture.}
    \label{fig:lha_scenario}
\end{figure}

\begin{figure}
    \centering
    \includegraphics[width=1.0\linewidth]{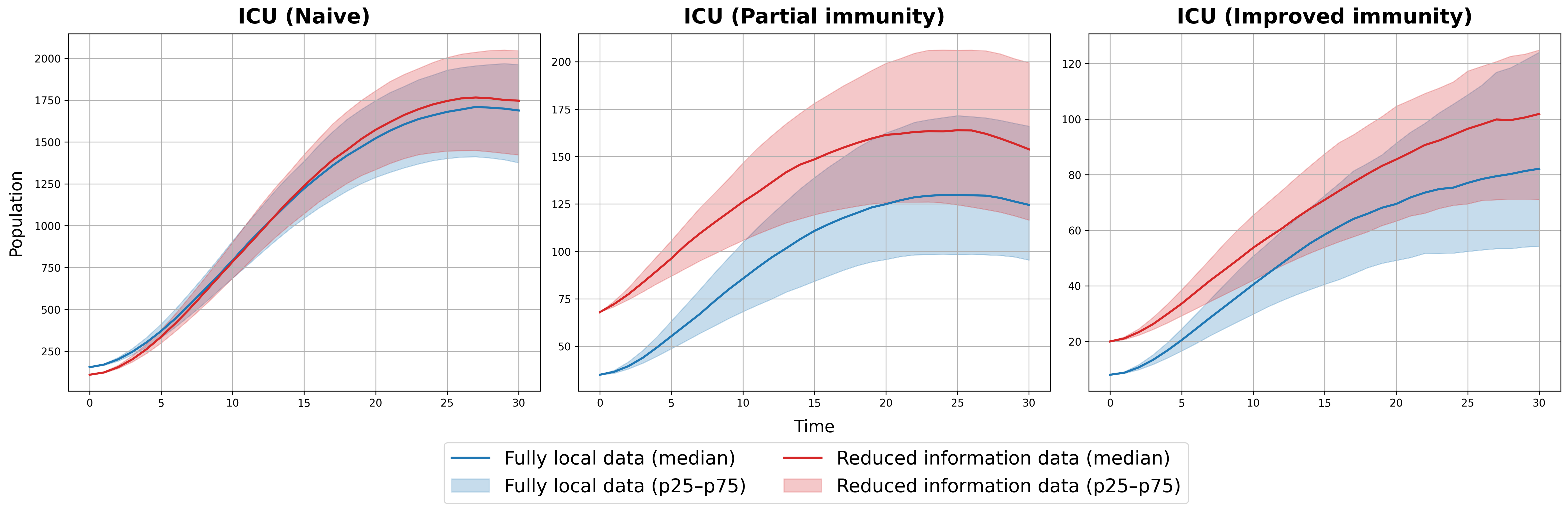}
    \caption{\textbf{Outcomes of two simulations with synthetic local and global data.} We show simulations with a steep increase in ICU numbers over 30 days only differing in initial conditions. Initial infections are either assigned with full local information to immunity and vaccination layers (blue) or, where immunity layers and infection data are decoupled on the global scale, conducted based on appropriate random weights (red).}
    \label{fig:lha_data_comparison}
\end{figure}

To test our platform's capability to integrate sensitive local data, we set up a proof-of-concept pipeline. In this pipeline, we generate synthetic local data through the MEmilio-ABM~\cite{KERKMANN2025110269} offering individual information containing infection state and vaccination histories. With this local data set we are mimicking data that can be uploaded to our platform by LHAs. In this synthetic local data, we have, for any symptomatic, severe, and critical infection, direct information on the number of prior vaccinations which, in a real-world setting, could be obtained locally from infected individuals. However, on the country-wide scale, vaccination is reported separately from infections and intensive care treatments~\cite{robert_koch_institut_2025_17623109}. For the demonstration here, the synthetic data generated for the city of Cologne (see~\cref{fig:lha_scenario}) is uploaded to our protected storage -- which is not accessible by the broader public or other LHAs. In order to show the advantage of local information integration, we then initialize the population either with the full information ("full local data") or by assigning infections, with appropriate weights, randomly ("reduced information data") to the three immunity layers (naive, partial, and improved); see~\cref{fig:lha_data_comparison}.

From~\cref{fig:lha_data_comparison}, we see that the original, local data set contains 155 critical infections in the naive immunity layer at the beginning of the simulation while the randomized initialization only assigns 110 critical infections to this layer. Then, during the simulation, the situation turns around, with 1688 critical infections for the local data set and 1746 critical infections for the data set with reduced information in the naive immunity layer . In total, while both simulations start with 198 critical infections, at the end of the simulation horizon the result based on the original data set yields 1894 critical infections versus 2002 in the reduced data set, an increase by 5.7~\%. While it has to be noted that the simulation represents a drastic increase of intensive care treatments in absolute numbers, the relative increase can be considered to be representative over many other scenarios and in cases where intensive care capacity is close to be exhausted a difference of 5-6~\% in patient numbers can become considerable.

\subsection{End-to-end runtime analysis}\label{sec:performance}

For a platform intended for epidemic or pandemic settings, performance is crucial. We therefore monitored the ESID pipeline as our core element for automated data ingestion, model preparation, scenario generation including simulation and data upload, in a preliminary version, over a representative evaluation period from 27~February to 10~March~2025. The goal was to analyze the runtime behavior of the complete workflow during regular operation. This evaluation targets the pipeline as an integrated system and does not aim to benchmark the underlying orchestration platform (i.e., Airflow running on Kubernetes). The analysis focuses on the stability and execution time of the necessary daily runs. In addition to the baseline Kubernetes-based execution, we analyzed the effect of offloading the computationally dominant task of the scenario generation phase to the JURECA HPC cluster. 

Each task in the pipeline defines its CPU and memory requirements using Kubernetes resource specifications. Typically, tasks related to data ingestion are lightweight and request 1~vCPU and 4~GB of memory, with upper limits of 4~vCPUs and 12~GB of memory. Tasks in the compute-intensive phases, namely model preparation and scenario generation, are assigned more resources with requests of 2~vCPUs and 32~GB of memory, and limits up to 8~vCPUs and 64~GB of memory. For offloading to HPC resources, we use a single JURECA compute node consisting of two AMD EPYC processors (2×64 cores) and 512\,GB of DDR4 memory.

\begin{figure}[!t]
    \centering
    \includegraphics[width=\textwidth]{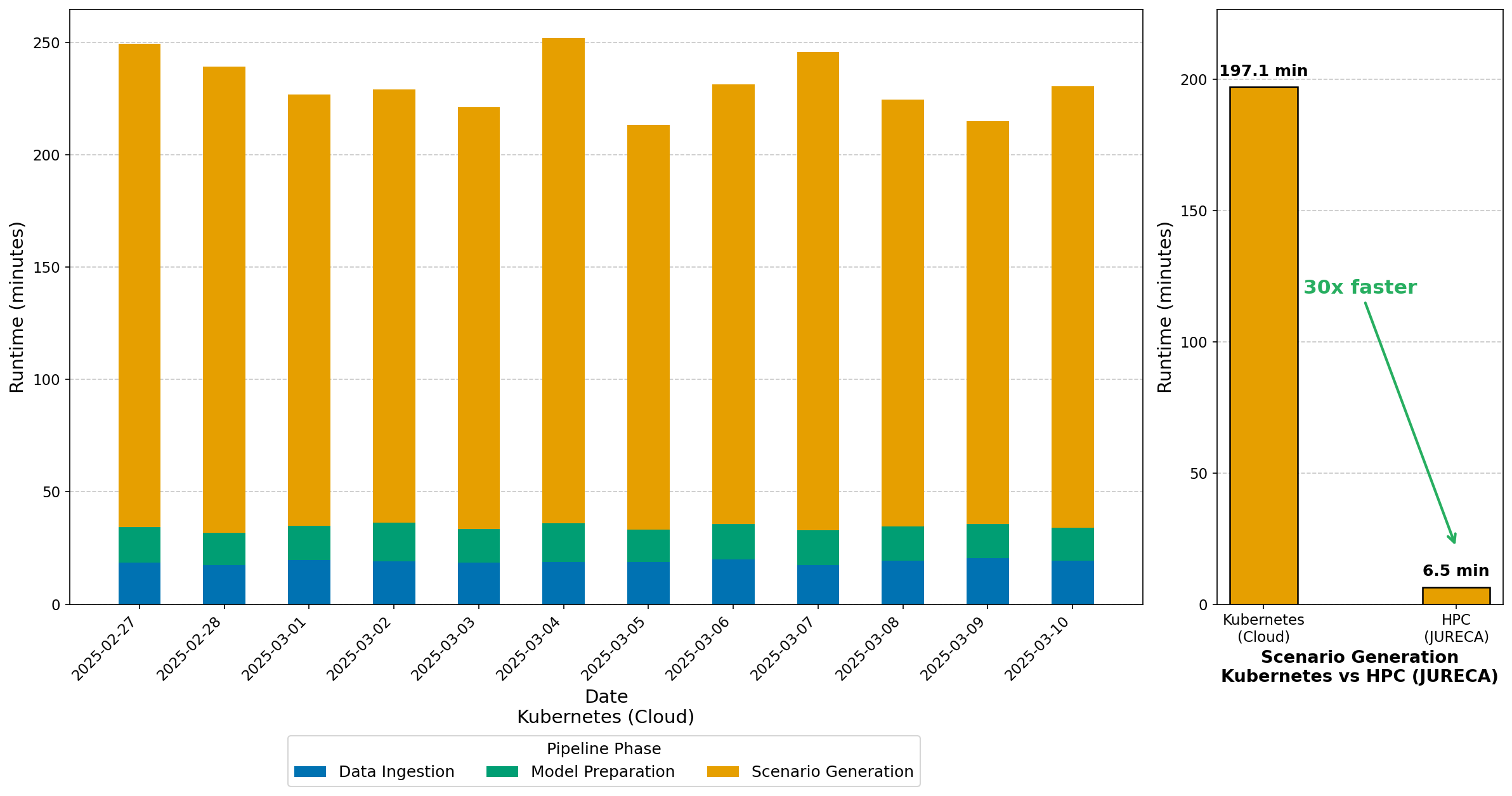}
    \caption{\textbf{Daily end-to-end runtime distribution of the ESID pipeline measured over the evaluation window (27~February to 10~March~2025).} On the left, results for Kubernetes-based execution are shown as a stacked breakdown by pipeline phases. On the right, a direct comparison of the scenario generation phase under Kubernetes and when offloaded to HPC resources via UNICORE on JURECA is shown.}
    \label{fig:stacked_runtime}
\end{figure}

\cref{fig:stacked_runtime} shows the daily end-to-end runtimes of the ESID pipeline over the evaluation period, presented as a stacked distribution across pipeline phases. Across the runs, the total execution time varies only slightly and stays within a narrow range, indicating stable behaviour at the DAG level. The runtime breakdown by phase shows that the scenario generation phase (represented by the \texttt{compute-scenarios} step) consumes most of the execution time, typically more than 80~\% of the total runtime. Data ingestion and model preparation tasks contribute only a small and stable fraction. 
The scenario generation phase is the most compute-intensive part of the pipeline, as it runs simulations for 33 scenarios on a coupled graph of 400 districts with ten runs each and subsequently uploads the results to the ESID back-end. When executed on a single HPC node, this phase achieves a 30-fold speedup, reducing the computation time from about 197~minutes to 6--7~minutes.

The evaluation shows that the pipeline can support workloads with varying requirements by leveraging both cloud and HPC resources, where the execution environment for each task is explicitly specified in the DAG definition.

\subsection{Testing and user experience}\label{sec:UX}

While technical checks and the provision of source code are minimal requirements for our platform to be reusable, we continuously tested our platform with the four pilot health authorities, from a user's perspective, to ensure its usefulness in a novel epidemic setting. An inclusive development process with a regular user-feedback loop and a rigorous user testing strategy was employed to ensure that the developed platform meets the requirements of all stakeholders and end-users. To do this, product requirements and user stories were collected from LHAs and further refined throughout the entire development process. 

As performance and immediate response are most critical for a useful web application, we minimized the size of the bundled scripts and assets. Libraries were carefully selected and optimized, with attention to dependency trees and tree-shaking, to keep the final bundle size as small as possible, currently resulting in a bundle size of just below 500 KB. Code splitting and lazy loading were employed to load certain functionalities only when needed. As not all functionality was required from the start, additional features are loaded on the fly. Using a consumer laptop, we measured the initial load time of the web front-end down for the first contentful paint as 0.8 seconds. Furthermore, we used Lighthouse~\cite{lighthouse}, a tool to objectively test multiple characteristics of websites, to evaluate page load time, accessibility, and best practices along a continuous integration (CI) workflow, when new features were added, or when triggered manually.
The Lighthouse performance score is overall 93~\%. This metric is derived as a weighted score of several common page load metrics, such as the time for the first contentful paint and the total blocking time until user interaction is possible. In the Accessibility audit, which tracks accessibility issues based on WCAG guidelines~\cite{wcag} with the page, the web application scores 90~\%. In the "Best Practices" section, our web application achieves 96~\%.

In addition, we evaluated the platform through its front-end in the form of a workshop to assess two main objectives: 1) the intuitiveness and user-friendliness of the user interface and 2) the added value that the ESID front-end offers to LHAs. Item (1) focused on the first impression and intuitive usability of the platform. Here, we evaluated the coherence of the UI components (map, scenarios, diagrams) and various design elements. Item (2) aimed to conduct an overall assessment of the added value of the platform for various tasks in health authorities – such as monitoring, evaluation, and decision support – as well as the satisfaction of LHAs with the platform.

\begin{figure}[!t] 
    \centering
    \includegraphics[width=\textwidth]{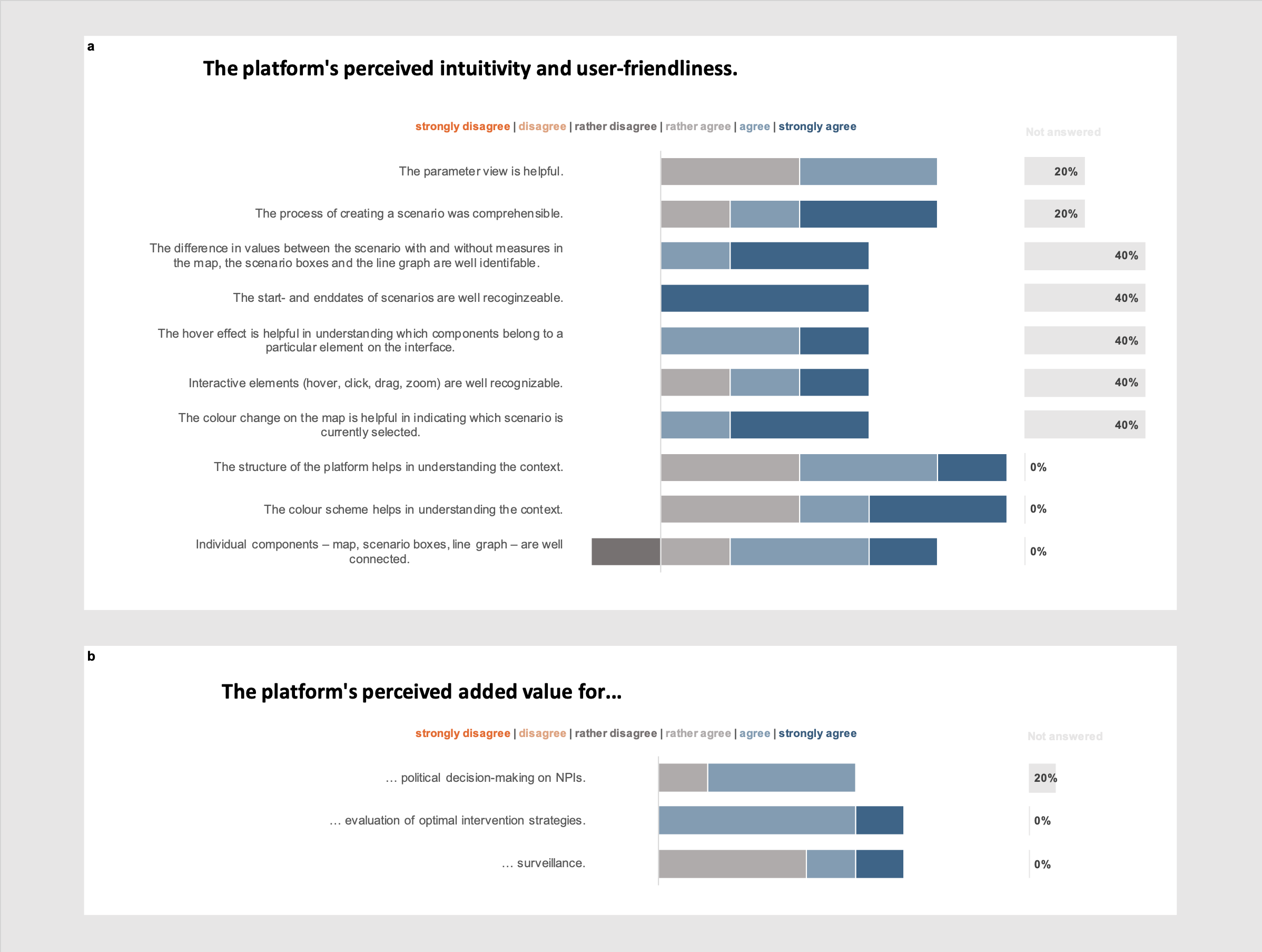}
    \caption{\textbf{Evaluation on intuitiveness, user-friendliness, and added value of the platform.} This figure shows the responses of test users from different LHAs (n=5) to questions about the intuitiveness and user-friendliness of the platform (Panel a) and its perceived added value (Panel b). The survey questions have been transformed into statements with respective participant's ratings on a 6-point Likert scale ranging from strongly disagree (orange) to strongly agree (blue). The results are provided in percentage, visually represented by bars of different colors and separate bars for "not answered". The size of the bars corresponds to the number of responses, i.e., the relative percentage within a category.} 
    \label{fig:Intuitivity_user-friendliness_added value}
\end{figure}
\begin{figure}[!t] 
    \centering
    \includegraphics[width=\textwidth]{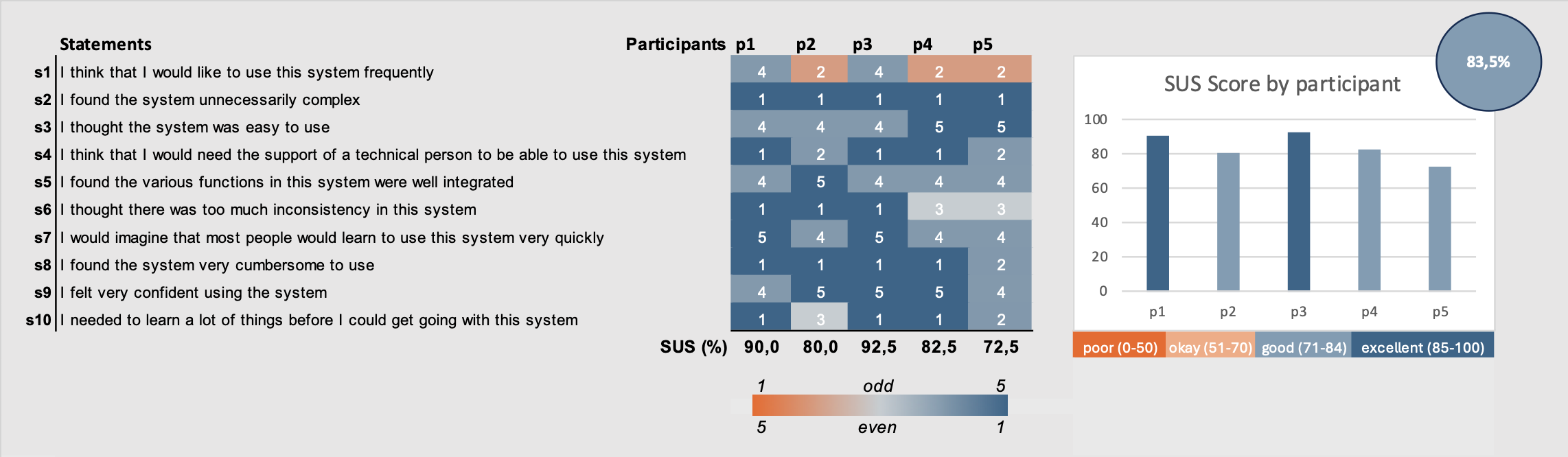}
    \caption{\textbf{System Usability Scale (SUS) score.} This figure shows the 10 statements of a standardized SUS evaluation addressing a system's ease of use, complexity, consistency, and learnability. The structure follows an alternating pattern of positive and negative statements that users rated on a 5-point Likert scale (from "strongly disagree" to ‘strongly agree"). 
    The heat map visualizes the ratings of 5 test users each representing a different LHA (p1-p5). The numbers in the heat map are the actual ratings that were provided. In order to calculate the SUS score 1 needs to be subtracted from the user's Likert ratings for odd-numbered statements and the the user's Likert ratings for even-numbered statements have to be subtracted from 5. Each item score will range from 0 to 4. The sum of the numbers multiplied by 2.5 will result in a score between 0 and 100. The heatmap's color scheme reflects the alternating structure of the statements in line with the color scheme of the SUS score categories in the bar chart  - from poor/dark orange to excellent/dark blue. Poor (0-50); okay (51-70); good (71-84); excellent (85-100). The bar chart depicts the individual SUS each test user scored, reaching an overall SUS score of 83.5~\%.} 
    \label{fig:SUS score}
\end{figure}

Through the qualitative evaluation of five data sets – as the workshop was conducted with a small group of LHA test users – a proof of concept was achieved. The results were collected from moderated feedback groups within the workshop and in the form of surveys on objectives (1) and (2). The collected ratings, which were based on multi-level scales, were used to create a series of selected key performance indicators (KPIs) -- Net Promoter Score (NPS), Customer Satisfaction Score (CSS), and System Usability Scale (SUS).

First, we asked test users to rate on a scale of 1 to 10 how likely they would recommend the platform to another LHA. Individual ratings showed a rather neutral attitude towards the platform, resulting overall in a Net Promoter Score (NPS) of -20. The NPS has a range of -100 to +100 calculated by subtracting the percentage of "detractors" (scores 0-6) from the percentage of "promoters" (scores 9-10). A higher score indicates greater customer loyalty and a stronger potential for growth. As overall promotion without testing in a real pandemic setting might be difficult to judge, we then evaluated more specific aspects with more targeted questions; see~\cref{fig:Intuitivity_user-friendliness_added value}. 

\cref{fig:Intuitivity_user-friendliness_added value}a) shows an overall good reception for multiple questions -- focusing on usability and intuitiveness such as the accessibility and connection of different components such as the maps, scenario boxes and the line graph, the platform's structure and color scheme, as well as interactive elements such as hover, click, drag and zoom. None of the test users had a negative impression or experienced difficulties understanding and using the platform. The user interface was rated as clearly structured and understandable. The relationship between the map view, scenario boxes and diagrams is comprehensible. Nearly 74~\% of user responses were positive (agree or strongly agree). The overall satisfaction of test users with the platform was collected in a Customer satisfaction Score (CSS). The CSS scored 80~\% satisfied test users. It was collected by asking users to rate their experience on a numerical scale from 1-6. To calculate the percentage the number of satisfied customers, i.e., users, was divided by the total number of responses, then multiplied by 100. 

In \cref{fig:Intuitivity_user-friendliness_added value}b), test users were asked to share their perception of the added value for routine tasks at LHAs, for which a continuous feedback loop throughout the project had identified an utility for the platform. The perceived added value of the platform in supporting LHAs in monitoring and early detection of outbreaks (surveillance) was fully recognized by one of the five test users. Another test user agreed that the platform may offer added value for surveillance, while three other rather agreed. In contrast, respondents are unanimously convinced that the platform offers significant advantages for assessing regional infection dynamics and evaluating the implementation of optimal intervention strategies, particularly the NPIs presented by the platform's scenarios. The function for displaying neighboring districts and the integration of local (regional) data were highlighted by LHAs as important features for regional risk management. Ultimately, the platform could be a useful tool for strengthening recommendations to decision makers with data-driven forecasts.

Finally, we applied the System usability Scale (SUS), a standardized method to measure the subjective usability of a system, such as the platform. A higher score indicates better perceived usability. The platform received a SUS score of 83.5~\%, within the upper level of the "good" category and close to excellent, as visualized by \cref{fig:SUS score}. The platform is generally perceived as intuitive and clear, particularly in terms of usability and visual feedback. The test users unanimously stated that they feel confident using the platform and can operate it without extensive prior knowledge or technical guidance.

\section{Conclusion}\label{sec:conclusion}

We presented a fully integrated software stack for epidemic and pandemic disease management with variable spatial resolutions, specific infection parameters and flexible intervention scenarios. 
The software components include data acquisition, transformation, ingestion to mathematical modeling and visualization leveraging a maximum of automation and HPC integration. We provided end-to-end runtime profiling with more than 30 scenarios on a large-scale model, demonstrating that our pipeline is a) capable of being run overnight on a daily basis or b) with the use of HPC resources in just several minutes. To the best of our knowledge, no such software stack for infectious disease forecasting workflows exists. We showed that with a high level of automation, large modeling tasks can be effectively accelerated to a level that makes it applicable in real world settings while avoiding manual interaction and efficiently using HPC resources. 

A corner stone of our project was the testing and evaluation of the user experience with our cohort of pilot LHAs and their employees to ensure a development that fits the users' needs in pandemic control and mitigation. We, thus, investigated
specific requirements such as the intuitive use to avoid substantial instruction delays or visualization and filtering along the multiple dimensions of infectious disease. 74~\% of user responses agreed or strongly agreed on user-friendliness and added value.

While we aimed at supporting public health experts and decision makers with mathematical modeling, our results satisfy requirements of a surveillance system. For the latter, a combination with early warning systems and indicators through wastewater monitoring would be beneficial to support outbreak detection. Then, daily automated pipelines could provide a well structured and spatially resolved view on ongoing disease dynamics and outbreaks.

The core components for mathematical modeling and visual analytics with the connecting REST API were designed and developed with a maximum of generality to allow for model adaptations to different spatial resolutions or courses of the disease. 
Thus, our publicly available open-source components can easily be adapted to different (administrative) units within the same country or transferred to other countries as long as corresponding data is available. More importantly, in the context of pandemic preparedness, the computational framework can be applied to other respiratory pathogens with pandemic potential.
This requires an adaptation of the disease model and front-end visualization by redefinition of disease parameters, groups and compartments, which is feasible because of the modular architecture of the framework.

\reviewanonymized{
\section{Acknowledgement}\label{sec:acknowledgement}

The authors gratefully acknowledge computing time on the supercomputer JURECA~\cite{JURECA} at Forschungszentrum J\"{u}lich under grant no.~\texttt{HPC4EPI}. The authors furthermore gratefully acknowledge the support of and exchange with Torge Korff, Annelene Kossow, Thomas Dau, Thies Marquardt, Susanne Krause from LHAs and Alexander Ullrich from the RKI.
The authors thank Uwe Naumann (RWTH Aachen University, i12 Software and Tools for Computational Engineering) for providing code for automatic differentiation.
This work was supported by the Initiative and Networking Fund of the Helmholtz Association (grant agreement number KA1-Co-08, Project LOKI-Pandemics)
and partially funded by the German Federal Ministry for Digital and Transport under grant agreement FKZ19F2211A and FKZ19F2211B (Project PANDEMOS).}

\section*{Software and data availability}

The mathematical models and scenario configurations are implemented in the MEmilio framework~\cite{bicker2026memilio}. The specific code is openly available at \url{https://github.com/SciCompMod/memilio/tree/esid-scenarios}.
The dagbags without secrets are available at \url{https://codebase.helmholtz.cloud/loki/public-dagbag}. The code for airflow to call is \url{https://codebase.helmholtz.cloud/loki/memiliflow}. The public docker image with our packages installed can be found at~\url{registry.hzdr.de/loki/memiliflow:main}.
The ESID front-end is available at \url{https://github.com/DLR-SC/ESID} and the back-end at \url{https://github.com/SciCompMod/ESID-backend}.
The Fleet/K8s deployment code is openly available at \url{https://codebase.helmholtz.cloud/loki/public-fleet-deployment-memilisql-restapi}.
The memilisql/data integration API code is openly available at \url{https://codebase.helmholtz.cloud/loki/public-memilisql-restapi}.

\bibliographystyle{ACM-Reference-Format}
\bibliography{literature.bib}
\end{document}